\documentclass[11pt]{article}
\usepackage{amsmath,amsfonts,amssymb,amsthm,amscd,accents,mathtools}
\usepackage{epsfig,url,cite,graphicx,graphics,subfigure,psfrag}
\usepackage{vmargin} 
\usepackage{fancyhdr}
\newcommand{\single}{\renewcommand{\baselinestretch}{1.05}\large\normalsize}

\newcommand{\B}[1]{\mathbf{#1}} 

\newcommand{\norm}[1]{\lVert#1\rVert} 
\newcommand{\sinc}[1]{\text{sinc}(#1)} 
\newcommand{\expect}{\mathbb{E}\,}
\newcommand{\real}{\mathbb{R}}
\renewcommand{\Re}{\text{Re}}
\renewcommand{\Im}{\text{Im}}
\newcommand{\Rank}{\text{rank}}
\newcommand{\var}[1]{\text{var}\big(#1\big)}
\newcommand{\covar}[1]{\text{covar}\big(#1\big)}
\newcommand{\uextend}{\widetilde{\widehat {\B{u}}}}
\newcommand{\tr}[1]{\text{tr}\big[#1\big]}
\newtheorem{theorem}{Corollary}

\setpapersize{A4}
\setmargins{1in}{.75in}{6.25in}{9in}{14pt}{10pt}{20pt}{30pt}
\title{\vspace*{-20pt}\textbf{Compressive and Noncompressive Power Spectral Density Estimation from Periodic Nonuniform Samples}}
\author{Michael~A.~Lexa, Mike~E.~Davies and John~S.~Thompson\\ \url{{michael.lexa, mike.davies, john.thompson}@ed.ac.uk}\\ \\
Institute of Digital Communications\\
University of Edinburgh}
\date{12 Oct 2011, Revised 15 Apr 2012}

\begin{document}
\sloppy\single
\maketitle

\begin{abstract}
This paper presents a novel power spectral density estimation technique for band-limited, wide-sense stationary signals from sub-Nyquist sampled data.
The technique employs multi-coset sampling and incorporates the advantages of compressed sensing (CS) when the power spectrum is sparse, but applies to sparse and nonsparse power spectra alike.
The estimates are consistent piecewise constant approximations whose resolutions (width of the piecewise constant segments) are controlled by the periodicity of the multi-coset sampling.
We show that compressive estimates exhibit better tradeoffs among the estimator's resolution, system complexity, and average sampling rate compared to their noncompressive counterparts. 
For suitable sampling patterns, noncompressive estimates are obtained as least squares solutions.
Because of the non-negativity of power spectra, compressive estimates can be computed by seeking non-negative least squares solutions (provided appropriate sampling patterns exist) instead of using standard CS recovery algorithms.
This flexibility suggests a reduction in computational overhead for systems estimating both sparse and nonsparse power spectra because one algorithm can be used to compute both compressive and noncompressive estimates. 
\end{abstract}

\section{Introduction}
Compressed sensing (CS) is a data acquisition strategy that exploits the sparsity or compressibility of a signal~\cite{candes_romberg_tao2006,candes_tao2006,donoho2006,candes_wakin2008}.
Typically, a signal is defined to be \emph{sparse} if its representation in an orthogonal basis contains only a few nonzero coefficients and is \emph{compressible} if it can be well approximated by a sparse signal.
In its simplest setting, CS proposes to sample a sparse discrete signal $\B{x}\in \real^{n}$ by acquiring $m<n$ inner product measurements $y_{l}=\langle \B{a}_{l},\B{x}\rangle$, $0\leq l \leq m$, where $\B{a}_{l}$ are the rows of a $m\times n$ measurement matrix $\B{A}$.
In matrix notation, $\B{y}=\B{A}\B{x}$.
Remarkably, CS asserts that $\B{x}$ can be accurately (if not exactly) reconstructed from the measurements $\B{y}$ even when the linear system of equations is underdetermined, provided $\B{A}$ satisfies certain conditions.
When CS is applied to the problem of sampling continuous-time signals, it has been shown that the undersampling of finite length vectors translates into sub-Nyquist sampling rates, where CS provides the theoretical justification and tools to reconstruct the signal despite the spectral aliasing that can occur~\cite{romberg2009,tropp_etal2010,mishali_eldar2010,lexa_davies_thompson_sp2012}.

In this paper, we study the problem of estimating the power spectral density (PSD) of a wide-sense stationary (WSS) random process within the context of a sub-Nyquist nonuniform periodic sampling strategy known as multi-coset (MC) sampling\cite{feng_bresler1996,pingfeng_phdthesis1997,venkataramani_bresler2000,bresler2008,mishali_eldar2009a,herley_wong1999,vaidyanathan_liu1990}.
Because of its nonuniform periodic nature, MC sampling is closely related to CS,%
\footnote{In fact, it can be formulated as a CS sampling strategy.}
and consequently, the proposed estimator easily incorporates and exploits CS theory.
The estimator however applies broadly, and in particular, applies to cases where the PSD is not sparse.
 
\begin{figure*}
\begin{minipage}{0.48\linewidth}
\subfigure[]{\centerline{\includegraphics[width=7cm]{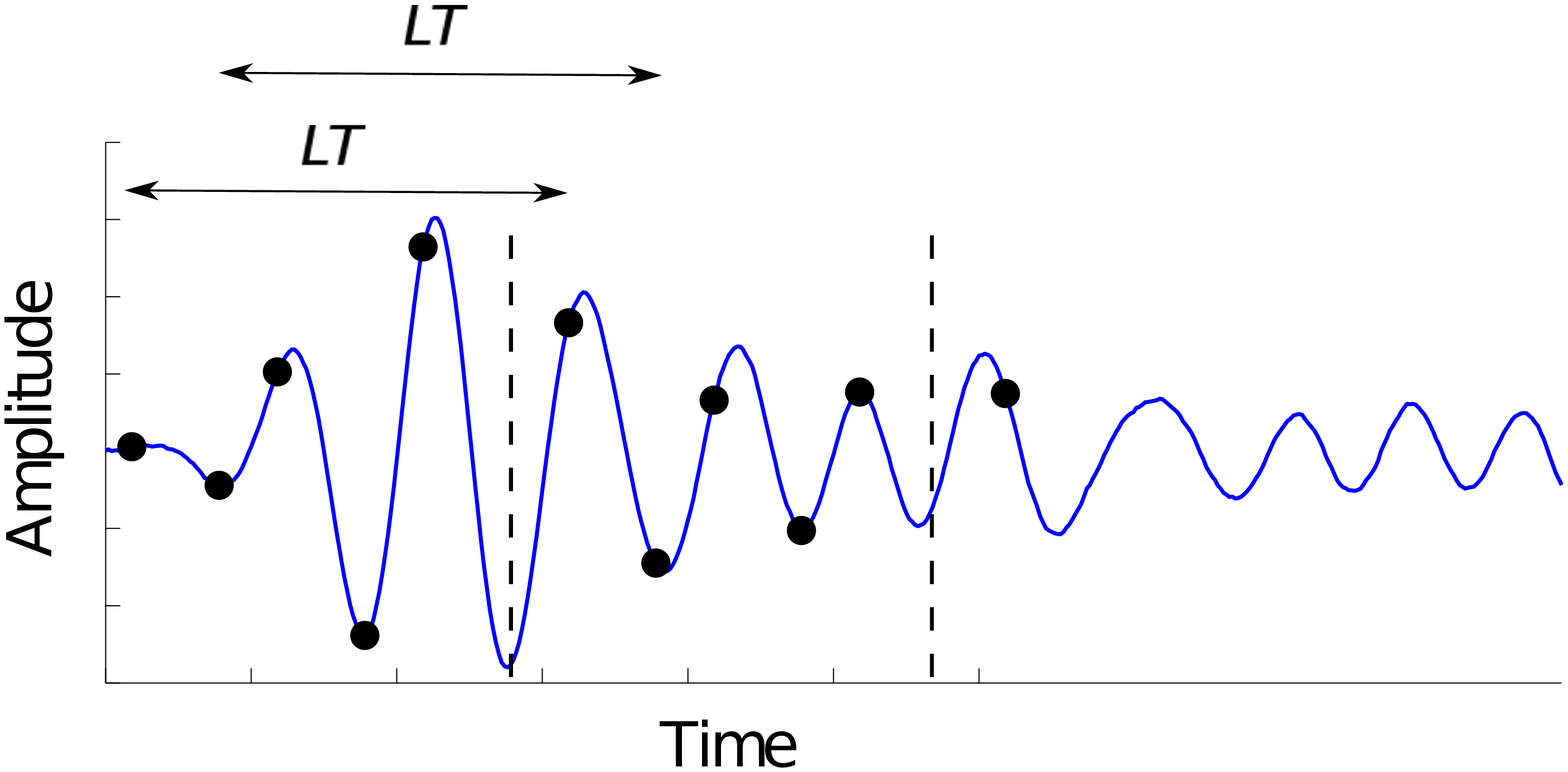}}}
\end{minipage}
\begin{minipage}{0.48\linewidth}
\subfigure[]{\centerline{\includegraphics[width=5cm]{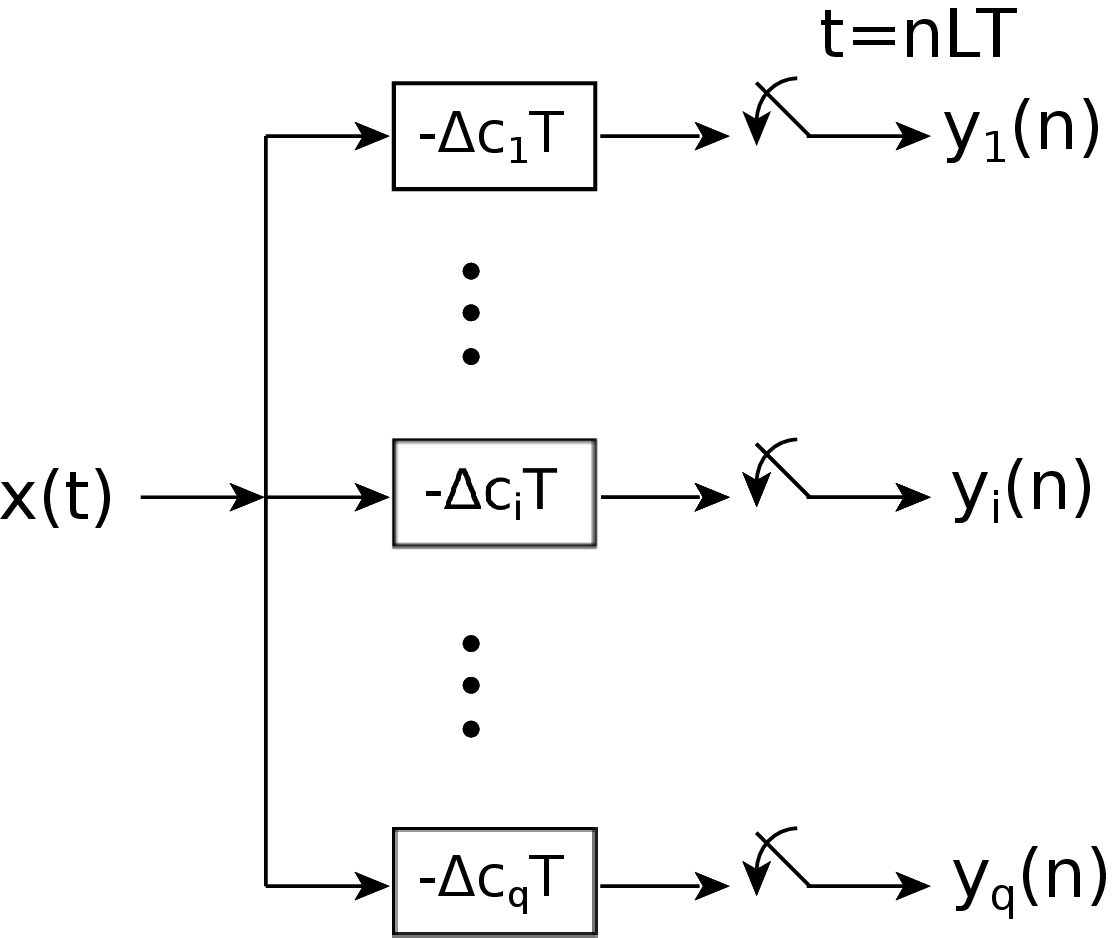}}}
\end{minipage}
\caption{(a) An example of multi-coset sampling. Samples (dots) are acquired at nonuniformly spaced points over $LT$ seconds with additional samples acquired at the same relative points in time over successive blocks of $LT$ seconds. (b) Multi-coset sampler implemented as a multi-channel system.}\label{fig:multicoset}
\end{figure*}

For a fixed time interval $T$ and for a suitable positive integer $L$, MC samplers sample an input signal $x(t)$ at the time instants $t=(nL+c_i)T$ for $1\leq i\leq q$, $n\in\mathbb{Z}$, where the time offsets $c_i$ are distinct, non-negative integers less than $L$ ($0\leq c_i< L$).
The strategy is depicted in Fig.~\ref{fig:multicoset}.
The sampling times are known collectively as the multi-coset \emph{sampling pattern} and the sets $\{(nL+c_i)T: n\in\mathbb{Z}, c_i\in\{0,\ldots,L-1\}\}$ are cosets of $\{nT:n\in\mathbb{Z}\}$.
Because $T$ is fixed throughout the paper, we simply refer to $\{c_{i}\}$ as the multi-coset sampling pattern instead of the actual sampling times.
MC samplers are parameterized by $q, L$, and $\{c_{i}\}$ and exhibit an average sampling rate of $q/LT$ Hz.
They are most easily implemented as multichannel systems as shown in Fig.~\ref{fig:multicoset} where channel $i$ shifts $x(t)$ by $c_{i}T$ seconds and then samples uniformly at $1/LT$ Hz.

Here, we assume a multichannel MC sampler and propose a novel nonparametric PSD estimator based on the MC samples.
The method estimates the average power within specific subbands of a WSS random process.
Hence it produces piecewise constant estimates that are in contrast to those supported on a discrete frequency grid, e.g., the periodogram implemented using the discrete Fourier transform (DFT).
Moreover, the estimator does not use passband filters to isolate subband signal components.
Rather the estimator uses the spectral aliasing that occurs in each channel to underpin the formation of a linear system of equations whose solution is the PSD estimate of interest (see Section~\ref{sect:mc_psd} for details).

We categorize the solutions to this linear system as either \emph{compressive} or \emph{noncompressive} depending upon whether CS is employed in the recovery of the estimate.
Compressive estimates result from the application of CS theory and pertain to situations where the underlying power spectrum is sparse and the linear system of equations is underdetermined.
Noncompressive estimates pertain to cases where the linear system is overdetermined (or determined) with the power spectrum's sparsity being irrelevant.
In both cases, the uniqueness of the recovered estimates depends on the choice of the sampling pattern $\{c_i\}$ (not all sampling patterns lead to unique solutions).
Because of the special structure of the linear system, we show that in some cases compressive estimates can be computed using a non-negative least squares (NNLS) approach.
This result is advantageous because in these cases it eliminates the necessity of using two separate algorithms to compute compressive and noncompressive estimates.
(Note noncompressive estimates are normally computed as least squares solutions but can also be computed using NNLS; see Section~\ref{subsect:sol_noncomp} for details).

The proposed method also indicates the regions within the $(L,q)$ parameter space that allow for compressive and noncompressive estimates, and it illustrates the associated tradeoffs among quantities related to these parameters such as the estimator's resolution or the number of MC channels.
By comparing these tradeoffs one can gauge the benefit of appling CS theory in cases where the PSD is sparse.  
Perhaps surprisingly, the advantage of a compressive versus a noncompressive estimate is not simply an improved sampling rate, but rather improved tradeoffs~\cite{lexa_davies_thompson_spars2011}.
For example, for a fixed spectral resolution, a compressive estimate might achieve the same accuracy as a noncompressive estimate but with a smaller number of channels.
In addition, we show that both compressive and noncompressive estimates are statistically consistent and show that the noncompressive estimate is asymptotically efficient.

\subsection{Related work}
There is a growing literature in applying CS and other sub-Nyquist techniques to power spectrum estimation, most notably in the field of cognitive radios where there is a need to find underutilized bandwidth in a crowded spectrum~\cite{tian_giannakis2007,leus_ariananda2011,ariananda_leus2011,ariananda_leus_tian2011,stoica_babu_li2011a,stoica_babu_li2011b} 
Most pertinent to the present discussion, however, is the work of Ariananda, Leus and Tian~\cite{ariananda_leus2011,ariananda_leus_tian2011,leus_ariananda2011}.
These papers also address the problem of PSD estimation from sub-Nyquist samples specifically incorporating CS-based sampling systems (including MC sampling).
Notwithstanding similar goals and approaches, the estimator proposed here differs in two significant aspects.
First, their method estimates the cross-correlations of the MC sample sequences and then recovers an estimate of the auto-correlation sequence of the input random process (from which they form a periodogram estimate of the PSD).
In contrast, the method described here directly recovers a piecewise constant estimate from the MC cross-correlation estimates.
The main difference being that Araiananda et al. estimate \emph{samples} of the PSD whereas we estimate the \emph{average power within subbands}.
An implication of this difference is that the method of Ariananda et al. is more computationally expensive:
their method requires the formation of a large $q^2 \times L^2$ matrix in the construction of their $q^2 \times (2L-1)$ measurement matrix; in contrast, we do not require the construction of this large matrix for our smaller $q(q-1)/2+1 \times L$ measurement matrix.  
Second, we treat the compressive case and address the statistical nature of our estimator (specifically taking into account the finite availability of data), while Ariananda et al. do not.

\subsection{Main contribution}
In short, the paper's main contribution is the development of a new nonparametric, statistically consistent, piecewise constant PSD estimator based on sub-Nyquist MC samples.
When the underlying PSD is sparse, it seemlessly incorporates CS theory, letting the resulting compressive estimator exhibit better parametric tradeoffs compared to their noncompressive counterparts.
Moreover, when a suitable sampling pattern exists, a NNLS approach can be used to compute compressive estimates, suggesting a possible savings in computational overhead for systems estimating sparse and nonsparse power spectra.

\section{A PSD approximation from multi-coset samples}\label{sect:mc_psd}
In this section, we derive a piecewise constant PSD approximation assuming full knowledge of the MC sequences.
This approximation serves as the motivation and model for the noncompressive and compressive estimators derived in Sections~\ref{sect:psd_est}.
In Section~\ref{sect:ex}, we present three examples illustrating the character and properties of the estimators.

\subsection{PSD approximation}
Let $x(t)$ be a real, zero-mean, WSS random process with autocorrelation function ${r_{xx}(\tau)\triangleq \expect x(t_1)x(t_2)}$, $\tau=t_1-t_2$.
The Fourier transform (FT) $P_{xx}(\omega)$ of $r_{xx}(\tau)$ is the PSD of $x(t)$~\cite{papoulis1991}.
\begin{equation*}
r_{xx}(\tau) \xleftrightarrow{\hspace*{7pt}\text{FT\hspace*{7pt}}} P_{xx}(\omega)
\end{equation*}
The PSD quantifies the power in any given spectral band of $x(t)$ and is a second order description of the random process.
Because $r_{xx}(\tau)$ is real and symmetric, $P_{xx}(\omega)$ is also real and symmetric.
Throughout the paper, we assume $x(t)$ is a band-limited process, meaning that $P_{xx}(\omega)$ vanishes for $\lvert \omega \rvert >\pi W$ rad/s~\cite[p. 376]{papoulis1991}.
For the MC sampling, we set $T=1/W$ thereby referencing the average system sampling rate $qW/L$ to the Nyquist rate $W$.

To estimate $P_{xx}(\omega)$ using MC samples, we examine the cross-correlations of the output sequences $y_i(n)\triangleq x(n\tfrac{L}{W}+\tfrac{c_i}{W}), n\in\mathbb{Z}$.
Let $a$ and $b$ denote two channels indices, then
\begin{equation}\begin{split}\label{equ:theo_psd0}
r_{y_ay_b}(n,m)&=\expect y_a(n) y_b(m) \\
&= \expect x\big(n\tfrac{L}{W}+\tfrac{c_a}{W}\big) x\big(m\tfrac{L}{W}+\tfrac{c_b}{W}\big) \\
&= r_{xx}\big(\tfrac{L}{W}(n-m)+\tfrac{1}{W}(c_a-c_b)\big) \\
&= r_{xx}\big(\tfrac{L}{W}k+\tfrac{1}{W}(c_a-c_b)\big) \\
&= r_{y_ay_b}(k)
\end{split}
\end{equation}
where the third step follows from the wide-sense stationarity of $x(t)$ and the substitution $k=n-m$.
From \eqref{equ:theo_psd0}, we see that the cross-correlation function $r_{y_ay_b}(k)$ is equivalent to shifting the auto-correlation function $r_{xx}(\tau)$ by $\tfrac{1}{W}(c_a-c_b)$ and then uniformly sampling it at $W/L$ Hz.
These operations imply the following discrete time Fourier transform (DTFT) pair for $k\in\mathbb{Z}$ and $\omega\in\real$,
\begin{align}\begin{split}\label{equ:theo_psd1}
r_{y_ay_b}(k)&=r_{xx}\big(\tfrac{L}{W}k+\tfrac{1}{W}(c_a-c_b)\big) \\
&\updownarrow{\hspace*{2pt}\text{\scriptsize DTFT}} \\
\frac{W}{L}\sum_{m=\bigl\lceil \tfrac{L}{2} \bigl(\tfrac{\omega}{\pi W}-1\bigr)\bigr\rceil}^{\bigl\lfloor \tfrac{L}{2} \bigl(\tfrac{\omega}{\pi W}+1\bigr)\bigr\rfloor} &P_{xx}(\omega-2\pi\tfrac{W}{L}m)e^{j\tfrac{1}{W}(c_a-c_b)(\omega-2\pi\tfrac{W}{L}m)}
\end{split}
\end{align}
where the relation follows from the shifting and sampling properties of the DTFT~\cite[p. 98, 117]{roberts_mullisDSP}.
The summation limits are finite for a given $\omega$ because $x(t)$ is assumed band-limited.
(Here we also assume $L$ is even.)

\begin{figure}
\centerline{\includegraphics[width=6.5cm]{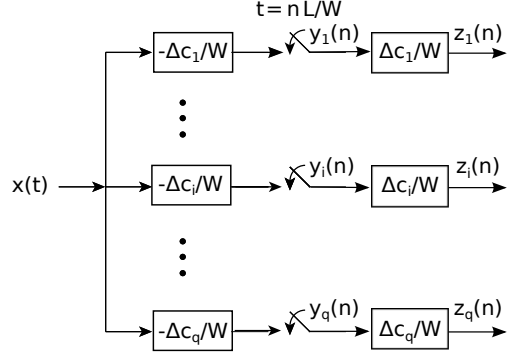}}
\caption{Block diagram of the multi-coset sampling system that shows the temporal shifting before and after sampling.}\label{fig:postprocessing}
\end{figure}

The phase shift $e^{j\tfrac{1}{W}(c_a-c_b)\omega}$ in~\eqref{equ:theo_psd1} arises from the time shifts introduced in the individual channels.
For our purposes, however, this phase shift needs to be removed.
This can be accomplished by shifting the sequences $y_i(n)$ by an amount equal to, but opposite, the initial delays (see Fig.~\ref{fig:postprocessing}).
Mathematically, this step can be expressed as
\begin{equation} \label{equ:theo_psd6}
z_{i}(n)\triangleq (y_{i}\star h_{i})(n), \quad n\in\mathbb{Z},\quad  i=1,\dotsc q
\end{equation} 
where $\star$ denotes convolution and $h_{i}(k)=\text{sinc}(\pi(k-c_i/W)), i=1,\dotsc q$, are the impulse responses of ideal fractional delay filters~\cite{Laakso_etal1996}.   
These filters are ``fractional'' in the sense that the time shifts are a fraction of the sampling period $L/W$.
For simplicity, we assume ideal fractional delay filters throughout the paper.
We note however that there is a large literature concerning the design and analysis of digital fractional delay filters (see~\cite{Laakso_etal1996} and the references wherein) ranging from highly accurate and computationally expensive filters to less accurate and computationally inexpensive ones.
The choice of design is largely application dependent and a detailed analysis of the impact of imperfect fractional delay filters and is beyond the scope of this paper.

By examining the cross-correlation functions $r_{z_az_b}(k)$, we see that the fractional delays indeed delay $r_{y_ay_b}(k)$ by $(c_a-c_b)/W$ seconds.
\begin{align}
r_{z_az_b}(k)&=\expect z_a(k+n) z_b(n) \nonumber \\
&= \sum_m \sum_l h_a(m) h_b(l) r_{y_ay_b}(k-m+l) \nonumber \\
&= \sum_{\alpha} \biggl[ \sum_m h_a(m) r_{y_ay_b}(\alpha-m) \biggr] h_b(\alpha-k) \label{equ:biasref} \\
&= r_{y_ay_b}(k) \star h_a(k) \star h_b(-k) \nonumber \\
&= r_{y_ay_b}(k) \star h_{a-b}(k) \label{equ:corr_delay}
\end{align}
where the second step follows from~\eqref{equ:theo_psd6}, $\alpha=k+l$, and $h_{a-b}(k)$ denotes the impulse of an ideal fractional delay filter with delay $(c_a-c_b)/W$ seconds. 

Equation~\eqref{equ:corr_delay} also implies that the DTFT of  $r_{z_az_b}(k)$ is the DTFT of $r_{y_ay_b}(k)$ multiplied by $e^{-j\tfrac{1}{W}(c_a-c_b)\omega}$.
Thus, by multiplying the frequency domain component in~\eqref{equ:theo_psd1} by $e^{-j\tfrac{1}{W}(c_a-c_b)\omega}$, we obtain the following \emph{inverse} DTFT relation 
\begin{equation}\begin{split}\label{equ:theo_psd2}
r_{z_az_b}(k)=&\frac{1}{2\pi}\sum_{m=-L/2+1}^{L/2} e^{-j\tfrac{2\pi}{L}(c_a-c_b)m}  \\ 
& \quad \int_{-\pi W/L}^{\pi W/L} P_{xx}(\omega-2\pi\tfrac{W}{L}m)\, e^{jk\tfrac{L}{W}\omega}~d\omega.
\end{split}
\end{equation}
Evaluating both sides of this equation at $k=0$ produces a linear system of equations over the set of all pairs $(a,b)$ of channels:
\begin{equation}\begin{split}\label{equ:theo_psd3}
r_{z_az_b}(0)=&\sum_{m=-L/2+1}^{L/2} e^{-j\tfrac{2\pi}{L}(c_a-c_b)m} \\
&  \qquad \underset{\displaystyle \triangleq P_{xx}(m)}{\underbrace{\int_{-\pi W/L}^{\pi W/L} P_{xx}\bigl(\omega-2\pi\tfrac{W}{L}m\bigr)~\frac{d\omega}{2\pi}}}.
\end{split}
\end{equation}
This linear system is the basis for our estimator.
For a given $m$, $P_{xx}(m)$ equals the average power of the process $x(t)$ in the spectral subband $[(2m-1)\pi W/L, (2m+1)\pi W/L]$.
Hence the set $\{\tfrac{L}{W} P_{xx}(m)\}$ forms a piecewise constant approximation to $P_{xx}(\omega)$.
Its resolution equals the width of the piecewise constant segments ($W/L$ Hz) and is inversely proportional to $L$, the parameter determining the period of the nonuniform samples.
Larger $L$ implies finer resolution; smaller $L$ implies coarser resolution.
The left hand side of~\eqref{equ:theo_psd3} is a weighted sum of the total average power of $x(t)$.
In Section~\ref{sect:psd_est}, we propose a statistical estimator based on this approximation that uses only a finite number of multi-coset samples.
Like the approximation $\{\tfrac{L}{W} P_{xx}(m)\}$, the estimator is piecewise constant and thus does not estimate PSD amplitudes at specific frequencies like standard parametric techniques or periodiogram estimates implemented using the DFT. 

Letting $i$ index the $\binom{q}{2}+1$ combinations%
\footnote{In actuality, the total number of pairs is $\binom{q}{2}+q$, but each of the $q$ cases where $a=b$ contributes a row of ones to $\B{\Psi}$. Thus to avoid the row dependency, only a single row of ones is included in $\B{\Psi}$.}
of pairs $(a,b)$, we can express~\eqref{equ:theo_psd3} in matrix notation,
\begin{equation}\label{equ:theo_psd4}
\B{u}=\B{\Psi} \B{v}
\end{equation} 
where $\B{u}\in\real^{q(q-1)/2+1}$, $\B{v}\in \real^{L}$ and
\begin{equation}\begin{split} \label{equ:theo_psd5}
&\B{u}=[u_{0},\dotsc,u_{q(q-1)/2}]^{T} ~\text{\footnotesize(T denotes transpose)} \\
&u_{i}=r_{z_az_b}(0)\\
&\B{\Psi}_{i,l}=e^{-j\tfrac{2\pi}{L}(c_a-c_b)_{i}m_{l}} \\
&\B{v}=[v_{0},\dotsc,v_{L-1}  ]^{T} \\
&v_{l}=P_{xx}(m_{l})
\end{split}
\end{equation}
for $i=0,\dotsc, q(q-1)/2$, $l=0,\dotsc,L-1$, and $m_{l}=-L/2+1+l$. 
The number of equations in this linear system can be doubled by separating $\B{\Psi}$ and $\B{u}$ into their real and imagery parts:
\begin{equation}\label{equ:theo_psd7}
 \underset{\displaystyle \triangleq \widetilde{\B{u}}}{\underbrace{\left[\begin{array}{c}
       \Re(\B{u})\\
       \Im(\B{u})
 \end{array}
\right]}}=
\underset{\displaystyle \triangleq \widetilde{\B{\Psi}}}{\underbrace{\left[ \begin{array}{c}
       \Re(\B{\Psi})\\
       \Im(\B{\Psi})
 \end{array}
 \right]}} \B{v}
 \end{equation}
where $\widetilde{\B{u}}$ and $\widetilde{\B{\Psi}}$ have dimensions $q(q-1)+1 \times 1$ and $q(q-1)+1 \times L$, respectively.

To calculate the PSD approximation, one would compute the set of cross-correlation values $\{r_{z_az_b}(0)=\expect z_a(n)z_b(n)\}$ for all pairs $(a,b)$ and solve~\eqref{equ:theo_psd7} for $\B{v}$.
Here we briefly consider two types of solutions, deferring a more detailed discussion until after the estimator is presented in Section~\ref{sect:psd_est} (see in particular Sections~\ref{subsect:sol_noncomp} and~\ref{subsect:sol_comp}).
First, we consider least squares solutions when $\widetilde{\B{\Psi}}$ is full rank (rank $L$) and~\eqref{equ:theo_psd7} is overdetermined.
Such solutions are referred to as \emph{noncompressive}.
Second, we consider CS solutions when~\eqref{equ:theo_psd7} is underdetermined and $\B{v}$ is sparse in the sense that only a few of its entries are nonzero.
These solutions are termed \emph{compressive}.

The existence of a unique least squares solution depends on the (column) rank of $\widetilde{\B{\Psi}}$ which in turn depends on the sampling pattern $\{c_i\}$.
In particular, a sampling pattern must be chosen such that $\widetilde{\B{\Psi}}$ has rank $L$ (implying that $q(q-1)+1\geq L$, or equivalently, that the number of rows of $\widetilde{\B{\Psi}}$ is greater than or equal to the number of columns). 
In Section~\ref{subsect:sol_noncomp}, we discuss two methods for constructing sampling patterns that produce full rank $\widetilde{\B{\Psi}}$, but given such a sampling pattern, the unique least squares solution is 
\begin{equation}\begin{split}
\B{v}_{NC}&\triangleq \arg\min_{\boldsymbol{\alpha}} \norm{\widetilde{\B{\Psi}}\boldsymbol{\alpha}-\widetilde{\B{u}}}_{2}^{2} \\
&= (\widetilde{\B{\Psi}}^{T}\widetilde{\B{\Psi}})^{-1}\widetilde{\B{\Psi}}^{T}\widetilde{\B{u}}
\end{split}\end{equation}
where $\norm{\cdot}_{2}$ denotes the $\ell_2$ norm, $(\widetilde{\B{\Psi}}^{T}\widetilde{\B{\Psi}})^{-1}\widetilde{\B{\Psi}}^{T}$ equals the pseudoinverse of $\widetilde{\B{\Psi}}$, and the subscript NC indicates that the approximation is noncompressive.
If $q(q-1)/2+1=L$ and $\widetilde{\B{\Psi}}$ is nonsingular, we simply have $\B{v}_{NC}= \widetilde{\B{\Psi}}^{-1}\widetilde{\B{u}}$.

Now, suppose $P_{xx}(\omega)$ is spectrally sparse, i.e., suppose its support has Lebesgue measure that is small relative to the overall bandwidth.
In addition, suppose $q(q-1)+1<L$ such that~\eqref{equ:theo_psd7} is an underdetermined linear system of equations.
Then from a CS perspective, $\B{v}$ is the sparse vector of interest that is to be recovered from the linear measurements $\widetilde{\B{u}}$, and $\widetilde{\B{\Psi}}$ is a (subsampled Fourier) measurement matrix.
The vector $\B{v}$ is sparse because its entries represent the average power of $x(t)$ in the subbands $[(2m-1)\pi W/L, (2m+1)\pi W/L]$ and $P_{xx}(\omega)$ is presumed to be sparse.
In this situation, a host of CS recovery algorithms may be used to compute compressive approximations $\B{v}_{C}$ provided an appropriate sampling pattern and a sufficient number of measurements for a given level of sparsity (see e.g.,~\cite{candes_romberg_tao2006b,needell_tropp2008, blumensath_davies2009}).
For example, for a suitable sampling pattern, the convex optimization problem
 \begin{equation}\label{equ:psd_est15}
\min_{\boldsymbol{\alpha}} ~\norm{\boldsymbol{\alpha}}_{1} \text{~subject to~} \widetilde{\B{u}}=\widetilde{\B{\Psi}}\boldsymbol{\alpha},
 \end{equation}
would yield a compressive approximation if $\mathcal{O}(S \log^{4}{L})$ measurements were acquired~\cite{rudelson_vershynin2008}, where $\norm{\cdot}_{1}$ denotes the $\ell_1$ norm and $S$ denotes the number of nonzero entries in $\B{v}$.
In Section~\ref{subsect:sol_comp}, we further exploit the structure of this linear system and advocate an algorithm that does not require an explicit sparsity constraint in the recovery procedure.

\section{Proposed multi-coset PSD estimator}\label{sect:psd_est}
The above PSD approximation is predicated on full knowledge of the output sequences $y_i(n)$, or equivalently, on having access to an infinite amount of data (an infinite number of multi-coset samples).
In this section, we derive and characterize an estimator based on~\eqref{equ:theo_psd4} that explicitly accounts for the finite availability of data.

\subsection{PSD estimation}\label{subsect:psd_est}
Again, let $x(t)$ by a real, band-limited, zero-mean WSS random process with PSD $P_{xx}(\omega)$.
We observe $x(t)$ over a finite duration interval and hence model the observed signal $x^{D}(t)$ as a windowed version of $x(t)$:
\begin{equation}\label{equ:psd_est0}
x^{D}(t)\triangleq x(t)w_R(t) 
\end{equation}
where
\begin{equation}
w_{R}(t)=\begin{cases}1 \qquad 0\leq t \leq D \\
0 \qquad \text{otherwise}. \end{cases}
\end{equation}
The MC samples of the windowed process $x^{D}(t)$ are
\begin{align}\begin{split}\label{equ:psd_est1}
y_{i}^{N}(n)&\triangleq x^{D}(\tfrac{1}{W}(nL+c_i)) \\
&=x(\tfrac{1}{W}(nL+c_i))\,w_{R}(\tfrac{1}{W}(nL+c_i)), 
\end{split}\end{align}
for $n=0,\dotsc,N-1$ and $i=1,\dotsc, q$, where $N=DW/L$ equals the number of samples acquired per channel during the $D$ second observation interval. 
(Here $N$ is assumed to be an integer.)
The fractionally delayed samples are (c.f.~\eqref{equ:theo_psd6})
\begin{equation}\label{equ:psd_est2}
z_{i}^{N}(n)\triangleq (y_{i}^{N}\star h_{i})(n)  
\end{equation}
where again $N$ signifies the number of samples%
\footnote{A perfect fractional delay filter has an infinite impulse response $h_i(n)=\text{sinc}(\pi(n-c_i/W))$ and thus the sequence $z_i^{N}(n)$ has, strictly speaking, infinite duration. However, in~\eqref{equ:psd_est2} we only consider $N$ output samples $z_i^{N}(n)$, ${n=0,1,\dotsc,N-1}$. We do not include another windowing operation because $z_i^{N}(n)$ will be finite for any practical, realizable filter.}.

Let $a$ and $b$ denote the indices of two channels. 
Then, given $z_{a}^{N}(n)$ and $z_{b}^{N}(n)$, the cross-correlation function $r_{z_az_b}(k)$ can be estimated by the sample correlation
\begin{equation}\label{equ:psd_est3}
\widehat{r}_{z_{a}z_{b}}^{\,N}(k)=\frac{1}{N} \sum_{n=0}^{N-\lvert k \rvert-1} z_{a}^{N}(k+n)z_{b}^{N}(n),
\end{equation}
for $0\leq \lvert k \rvert \leq N-1$.
Mimicking the standard derivation of a periodogram (see e.g.~\cite[p.321]{johnson_dudgeon1993}), we find the DTFT of $\widehat{r}_{z_{a}z_{b}}^{\,N}(k)$ to be
\begin{equation}\begin{split}\label{equ:psd_est4}
\widehat{r}_{z_{a}z_{b}}^{\,N}(k) \xleftrightarrow{\hspace*{5pt}\text{DTFT\hspace*{5pt}}} \frac{1}{N} \Big[Z_a^{N}(e^{j\omega \tfrac{L}{W}}) \Big] \Big[Z_b^{N}(e^{j\omega \tfrac{L}{W}})\Big]^{*}
\end{split}\end{equation}
where $Z_{a}^{N}(e^{j\omega \tfrac{L}{W}})$ and $Z_{b}^{N}(e^{j\omega \tfrac{L}{W}})$ are the DTFT of $z_{a}^{N}(n)$ and $z_{b}^{N}(n)$ respectively.
The transform $Z_{a}^{N}(e^{j\omega \tfrac{L}{W}})$ may be further decomposed in terms of $Y_{a}^{N}(e^{j\omega \tfrac{L}{W}})$ and $X^{D}(\omega)$ where
\begin{align}\begin{split}\label{equ:psd_est5}
x^{D}(t) &\xleftrightarrow{\hspace*{11pt}\text{FT\hspace*{11pt}}} X^{D}(\omega) \\
y_a^{N}(n) &\xleftrightarrow{\hspace*{5pt}\text{DTFT\hspace*{5pt}}} Y_{a}^{N}(e^{j\omega \tfrac{L}{W}}).
\end{split}\end{align}
Specifically, for $\omega\in \real$, we have from~\eqref{equ:psd_est1} and~\eqref{equ:psd_est2} that
\begin{align}\begin{split}\label{equ:psd_est6}
Z_{a}^{N}(e^{j\omega \tfrac{L}{W}})&=Y_{a}^{N}(e^{j\omega \tfrac{L}{W}})e^{-j\tfrac{c_a}{W}\omega}\\
&=\frac{W}{L} \sum_{m=-\infty}^{\infty} X^{D}(\omega-2\pi \tfrac{W}{L}m) e^{-j\tfrac{2\pi}{L}c_{a}m}.
\end{split}\end{align}

Because $x^{D}(t)$ is time limited, it cannot, strictly speaking, be band-limited.
However, for sufficiently long windows it is reasonable to assume that the vast majority of the energy of $x^{D}(t)$ lies within the band $(-\pi W, \pi W]$, i.e., it is reasonable to assume that with sufficiently long windows $x^{D}(t)$ is essentially band-limited to the same band as $x(t)$.
With this assumption, one period of $Z_{a}^{N}(e^{j\omega \tfrac{L}{W}})$ is well approximated by the finite summation,
\begin{equation}\label{equ:psd_est7}
Z_{a}^{N}(e^{j\omega \tfrac{L}{W}}) =\negthickspace \frac{W}{L}  \sum_{m=-\frac{L}{2}+1}^{\frac{L}{2}} X^{D}(\omega-2\pi \tfrac{W}{L}m) e^{-j\tfrac{2\pi}{L}c_{a}m}.
\end{equation}

Substituting~\eqref{equ:psd_est7} for $Z_{a}^{N}(e^{j\omega \tfrac{L}{W}})$ in~\eqref{equ:psd_est4} (along with a corresponding expression for $Z_{b}^{N}(e^{j\omega \tfrac{L}{W}})$)  and evaluating the inverse DTFT at zero yields 
\begin{equation}\begin{split}\label{equ:psd_est9}
\widehat{r}_{z_az_b}^{\,N}(0) =  & \sum_{m} \sum_{n} e^{-j\tfrac{2\pi}{L}(c_a m-c_b n)} \\ 
& \int_{-\pi W/L}^{\pi W/L} \frac{1}{D} X_m^{D}(\omega) [X_n^{D}(\omega)]^{*}~\frac{d\omega}{2\pi},
\end{split}\end{equation}
where $X_m^{D}(\omega)$ is shorthand notation for $X^{D}(\omega-2\pi \tfrac{W}{L}m)$.

We observe that for $m=n$, the integrands, $\frac{1}{D}\big\lvert X_m^{D}(\omega) \big\rvert^{2}, m=-L/2+1,\dotsc,L/2$, are periodogram estimates of $L$ disjoint subbands of $P_{xx}(\omega)$, and the corresponding integrals
\begin{equation}\label{equ:psd_est9.5}
\widehat{P}_{xx}(m) \triangleq \int_{-\pi W/L}^{\pi W/L} \frac{1}{D} \big\lvert X_m^{D}(\omega) \big\rvert^{2}~\frac{d\omega}{2\pi}
\end{equation}
are estimates of the power within these subbands.
The set $\{\tfrac{L}{W}\widehat{P}_{xx}(m)\}$ thus constitutes a piecewise constant estimate of $P_{xx}(\omega)$ at resolution $W/L$.

In contrast, the cross terms in~\eqref{equ:psd_est9} (i.e., terms for which $m\neq n$) asymptotically approach zero as $D$, or equivalently $N$, grows large.
To see this, rewrite the integrand in~\eqref{equ:psd_est9} as
\begin{align}
\frac{1}{D} & X_m^{D}(\omega) [X_n^{D}(\omega)]^{*} \nonumber \\
&= \frac{1}{D} X^{D}(\omega-2\pi \tfrac{W}{L}m) [X^{D}(\omega-2\pi \tfrac{W}{L}n)]^{*} \nonumber\\
\begin{split}
&=  \int_{-\infty}^{\infty} W_R(\alpha) X(\omega-2\pi \tfrac{W}{L}m-\alpha)~\frac{d\alpha}{\sqrt{D}} \\
& \qquad \times \int_{-\infty}^{\infty} W_{R}^{*}(\beta) X^{*}(\omega-2\pi \tfrac{W}{L}n-\beta)~\frac{d\beta}{\sqrt{D}},
\end{split}\label{equ:psd_est10}
\end{align}
where $W_{R}(\omega)=D\,\sinc{\tfrac{D}{2}\omega}e^{-j\tfrac{D}{2}\omega}$ is the FT of the rectangular window $w_{R}(t)$, $X(\omega)$ is the FT of $x(t)$ appropriately defined (see~\cite[p.  416]{papoulis1991}), and 
\begin{equation*}
\sinc{\alpha}=\begin{cases} 1,& \text{if~}\alpha=0 \\
\sin{(\alpha)}/\alpha, &\text{otherwise} \end{cases}.
\end{equation*} 
As $D\rightarrow \infty$, $W_R(\omega)$ approaches a delta function~\cite[p. 280]{papoulis1962}; however because of the factors $1/\sqrt{D}$, the integrals approach zero as $D\rightarrow \infty$ for $m\neq n$.

Because of this fact, we choose to approximate $\widehat{r}_{z_az_b}^{\,N}(0)$ using only the terms $m=n$ in~\eqref{equ:psd_est9}
\begin{equation}\label{equ:psd_est11}
\widehat{r}_{z_az_b}^{\,N}(0) \approx \sum_{m} e^{-j\tfrac{2\pi}{L}(c_a-c_b)m} ~\widehat{P}_{xx}(m).
\end{equation}
Letting $i$ index all pairs $(a,b)$, we arrive at an empirical counterpart to~\eqref{equ:theo_psd4}
\begin{equation}\label{equ:psd_est12}
\widehat{\B{u}}=\B{\Psi} \widehat{\B{v}}
\end{equation} 
where 
\begin{equation}\begin{split}\label{equ:psd_est13}
&\widehat{\B{u}}=[\widehat{u}_{0},\dotsc,\widehat{u}_{q(q-1)/2}]^{T} \\
&\widehat{u}_{i}=\widehat{r}_{z_az_b}^{\,N}(0) \\
\B{\Psi}&_{i,l}=e^{-j\tfrac{2\pi}{L}(c_a-c_b)_{i}m_{l}} \\
&\widehat{\B{v}}=[\widehat{v}_{0},\dotsc,\widehat{v}_{L-1}  ]^{T} \\
&\widehat{v}_{l}\approx \widehat{P}_{xx}(m_{l})
\end{split}\end{equation}
for $i=0,\dotsc, q(q-1)/2$, $l=0,\dotsc,L-1$, and $m_{l}=-L/2+1+l$. 
Similar to~\eqref{equ:theo_psd7}, we can double the number of equations by separating the real and imagery parts of $\B{\Psi}$ and $\widehat{\B{u}}$:
\begin{equation}\label{equ:psd_est14}
\uextend= \widetilde{\B{\Psi}} \widehat{\B{v}}
\end{equation}
where $\uextend=[\Re(\widehat{\B{u}})~\Im(\widehat{\B{u}})]^{T}$ and $\widetilde{\B{\Psi}}\in \real^{q(q-1)+1\times L}$ is defined in~\eqref{equ:theo_psd7}.
\emph{In this equation, $\widehat{\B{v}}$ is the PSD estimator we wish to solve for.}

\subsection{Noncompressive solutions (overdetermined case)}\label{subsect:sol_noncomp}
A unique least squares solution to~\eqref{equ:psd_est14} exists if $\widetilde{\B{\Psi}}$ has full column rank, i.e., $\Rank(\widetilde{\B{\Psi}})=L$.
Full rank implies $q(q-1)+1\geq L$, or that the system is overdetermined (here we treat a determined system as a special case of the overdetermined one).
In this case, a noncompressive estimate can computed using the pseudoinverse
\begin{equation}\label{equ:noncomp_sol1}
\widehat{\B{v}}_{NC}\triangleq (\widetilde{\B{\Psi}}^{T}\widetilde{\B{\Psi}})^{-1}\widetilde{\B{\Psi}}^{T}\uextend,
\end{equation}
where $(\widetilde{\B{\Psi}}^{T}\widetilde{\B{\Psi}})^{-1}\widetilde{\B{\Psi}}^{T}$ is the pseudoinverse of $\widetilde{\B{\Psi}}$.
A solution of this type is \emph{noncompressive} because it does not rely on CS theory, and in particular, it does not require $\widehat{\B{v}}$ to be sparse.

For a fixed $L$, a full rank $\widetilde{\B{\Psi}}$ requires (i) choosing $q$ large enough such that $q(q-1)+1\geq L$ and (ii) choosing a sampling pattern such that $\Rank(\widetilde{\B{\Psi}})=L$.
The first condition is required since $q(q-1)+1<L$ implies $\Rank(\widetilde{\B{\Psi}})<q(q-1)+1$, regardless of the sampling pattern.
The second condition is necessary since there exists sampling patterns that lead to $\widetilde{\B{\Psi}}$ having a non-empty nullspace even if $q(q-1)+1\geq L$, meaning that the least squares solution is no longer unique.
Section~\ref{subsect:noncomp_tradeoffs} discusses the tradeoff between $q$ and $L$ imposed by the first condition assuming the second condition is met.
Here we address the second condition.

\begin{figure*}
\begin{minipage}{.48\linewidth}
\centerline{\includegraphics[width=7cm]{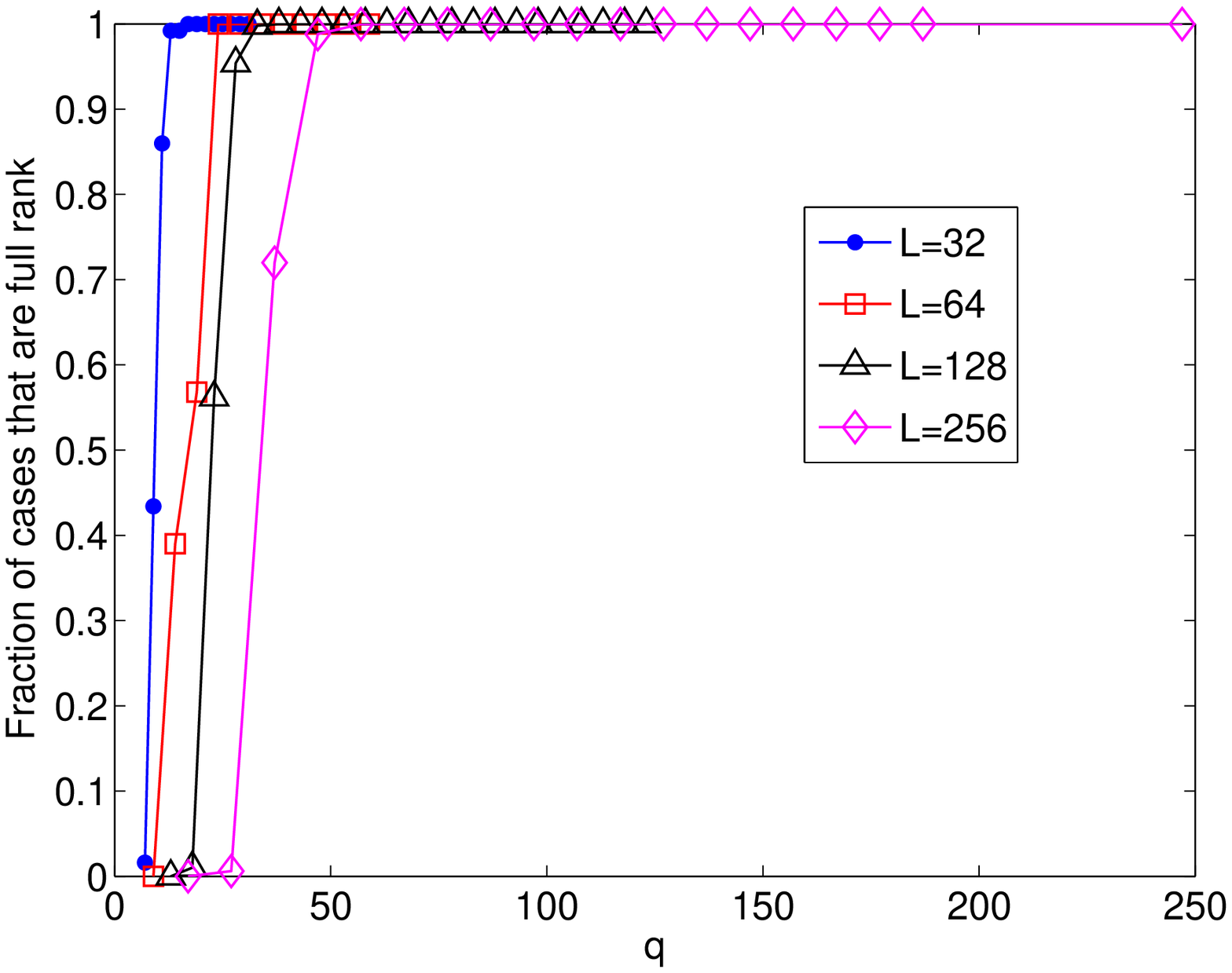}}
\end{minipage}\hfill
\begin{minipage}{.48\linewidth}
\centerline{\includegraphics[width=7cm]{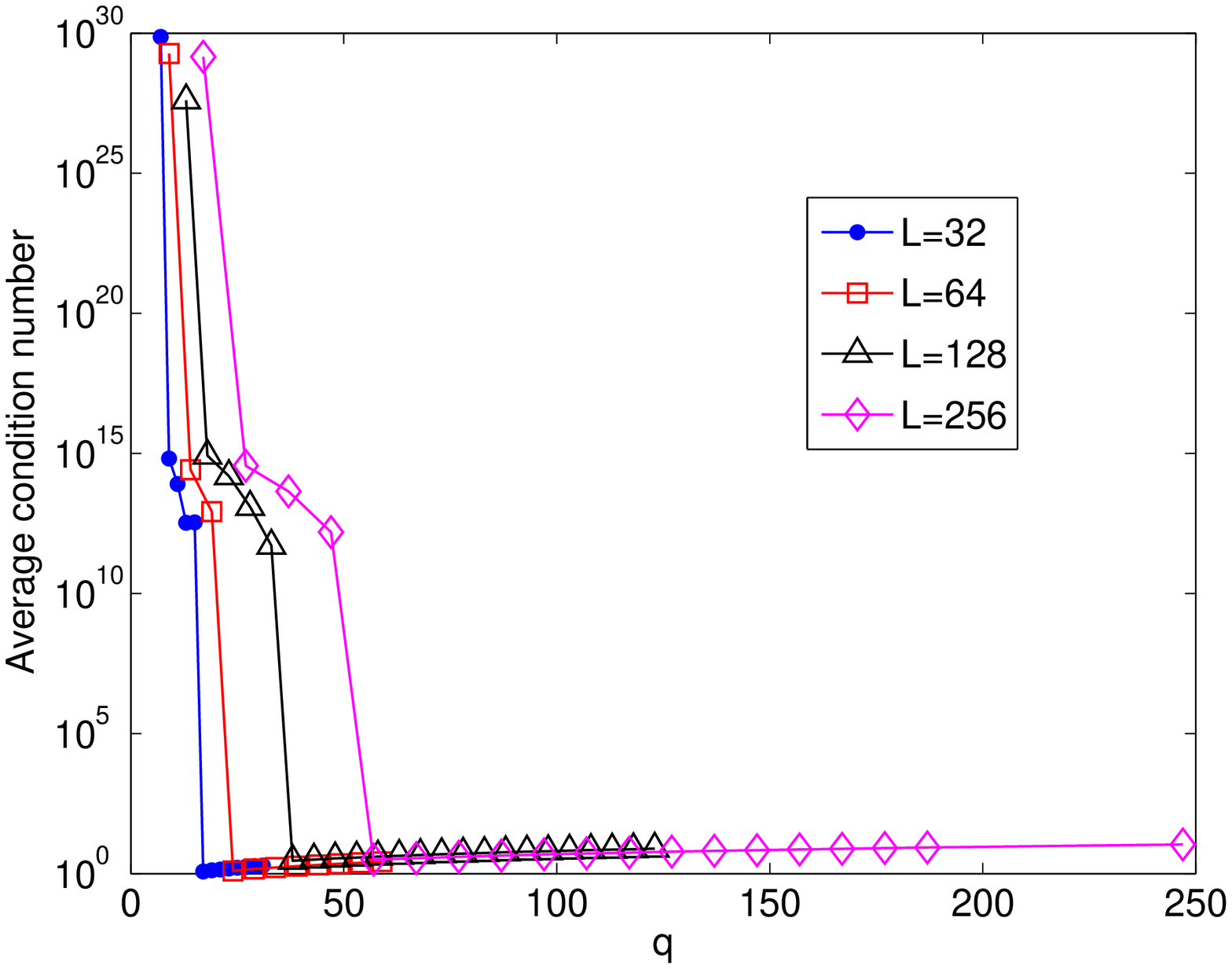}}
\end{minipage}
\caption{Four scenarios showing that random sampling patterns can produce well-conditioned, full rank $\widetilde{\B{\Psi}}$ after some threshold point in the noncompressive case. Left: Fraction of times random patterns produce full rank $\widetilde{\B{\Psi}}$ for different $(L,q)$ values. Right: Corresponding condition numbers.}\label{fig:rand_samp_pattern}
\end{figure*}

Fig.~\ref{fig:rand_samp_pattern} offers empirical evidence suggesting that generating a sampling pattern uniformly at random from $\{0,\dotsc,L-1\}$ can be a simple and effective way of constructing a full rank $\B{\widetilde{\Psi}}$, provided the ratio of its dimensions $\tfrac{L}{q(q-1)+1}$ is small enough. 
The plots show the results from four trials corresponding to four different values of $L$.
For each trial, $L$ was fixed and $q$ was allowed to range from its minimum value---the smallest integer value satisying $q(q-1)+1>L$---to its maximum value $L-1$.
For each $(L,q)$ pair, 500 sampling patterns (of length $q$) were generated and the fraction of those that produced full rank $\widetilde{\B{\Psi}}$ are recorded in the left hand plot.
The right hand plot displays the average condition number of  $\widetilde{\B{\Psi}}$ in each case.
Both plots suggest that after some threshold point random sampling patterns consistently produce well-conditioned and full rank matrices for the noncompressive problem.
For the four cases shown, the threshold occurs when the ratio $\tfrac{L}{q(q-1)+1}$ roughly equals 0.12, but a formal treatment of this phenomenon is beyond the scope of this paper. 

For larger $\tfrac{L}{q(q-1)+1}$ ratios (still falling within the noncompressive problem), full rank $\widetilde{\B{\Psi}}$ can sometimes be   constructed by choosing the sampling pattern to be a \emph{Golomb ruler}~\cite{drakakis2009}.
Golomb rulers are integer sets whose difference sets contain unique elements.
The cardinality of the integer set is the ruler's \emph{order} and the largest difference between two integers is its \emph{length}.
The idea is that the integers represent marks on an imaginary ruler and the elements of the difference set are all the lengths that can be measured by the ruler.  
The goal in the study of Golomb rulers is to find the minimum length rulers for a given order.
In the present context, however, we are only interested in using already tabulated rulers.
For example, for $L=64$ and $q=10$, one can take the minimum-length order-10 ruler to be the sampling pattern 
\begin{equation*}
\{c_i\}_{i=1}^{10}=\{1, 2, 7, 11, 24, 27, 35, 42, 54, 56\}.
\end{equation*}
Note that the ruler's length, 55, is less than $L=64$ as it should be for MC sampling.
This pattern yields a full rank $\widetilde{\B{\Psi}}$ with condition number $1.4$ whereas in Fig.~\ref{fig:rand_samp_pattern} random patterns \emph{never} produced a full rank matrix out of 500 test cases.
The implication of this improvement is somewhat limited however because there are relatively few known (up to order 26) minimum length Golomb rulers.
For instance with $L=1024$, the minimum $q$ value is 33, but there are no known minimum length Golomb rulers with orders greater than 26.
When applicable, Golomb ruler sampling patterns can be an effective means to obtain a full rank $\B{\Psi}$ when the ratio $\tfrac{L}{q(q-1)+1}$ is near the threshold value.
Note also that Ariananda et al. advocate the use of Golomb rulers in their study of multi-coset PSD estimation, but they used it to ensure the rank of a binary matrix~\cite{ariananda_leus_tian2011}.

To summarize the foregoing, one computes a noncompressive estimate by fractionally delaying each channel of MC samples $y_{i}^{N}(n), i=1,\dotsc,q$, computing the cross-correlation values $\{\widehat{r}_{z_az_b}^{\,N}(0)\}$, and solving~\eqref{equ:psd_est12} via~\eqref{equ:noncomp_sol1}.

\subsection{Compressive solutions (underdetermined and sparse case)}\label{subsect:sol_comp}
When~\eqref{equ:psd_est14} is underdetermined and the sampling pattern is chosen uniformly at random from $\{0,\dotsc,L-1\}$, $\widetilde{\B{\Psi}}$ is a randomly subsampled DFT matrix which is a well-known type of CS measurement matrix.
Thus a number of CS recovery algorithms (e.g.,~\cite{candes_romberg_tao2006b,needell_tropp2008, blumensath_davies2009}) will solve~\eqref{equ:psd_est14}, provided $\widehat{\B{v}}$ is sparse and the number of CS measurements (length of $\uextend$) is sufficient.  
However, because~\eqref{equ:psd_est14} has special structure, the usual CS recovery algorithms may be replaced by a non-negative least squares approach.
This claim rests on the following result of Donoho and Tanner~\cite{donoho_tanner2010} (see also~\cite{donoho_tanner2008}):
\begin{theorem}[\cite{donoho_tanner2010}, Corollary 1.4 and 1.5]\label{thm:donoho_tanner}
Assume $d$ is even and for $n=0,1,2,\dotsc,L-1$, let $\B{A}\in \real^{d\times L}$ be the partial DFT matrix ($d<L$)
\begin{equation*}
\B{A}_{k,n}=\begin{cases} \cos{\Big(\frac{\pi kn}{L}\Big)} & k=0,2,4,\dotsc,d-2 \\
\sin{\Big(\frac{\pi (k+1)n}{L}\Big)} & k=1,3,5,\dotsc,d-1 
\end{cases}
\end{equation*}
Let $\B{x}$ be a non-negative $s$-sparse vector, i.e., let $\B{x}$ have at most $s$ nonzero positive entries.
Then given the linear measurements $\B{b}=\B{A}\B{x}$, any approach that solves the linear system and maintains nonnegativity will correctly recover the $s$-sparse solution $\B{x}$, provided $d\geq 2s$.
\end{theorem} 
The corollary says that when $\B{x}$ is $s$-sparse and non-negative, one needs only $2s$ Fourier measurements to uniquely recover it.
Moreover, the theorem states that $\B{x}$ can be recovered using \emph{any} algorithm that produces a nonnegative solution to $\B{A}\B{x}=\B{b}$.
The fact that a $L$ dimensional $s$-sparse vector $\B{x}$ can be recovered from only $d<L$ Fourier measurments is an archetypical CS result.
However, in contrast to typical CS recovery strategies, signal recovery in this case does not require explicit use of a sparsity constraint like $\ell_1$ minimization (see for example~\eqref{equ:psd_est15}).
The reason is that this result is a specific case of a more general phenomenon in the recovery of sparse nonnegative vectors for underdetermined linear system of equations. 
In essence, it is known that if the row span of a measurement matrix intersects the positive orthant and if the vector to be recovered is nonnegative then the solution to the underdetermined linear system is a singleton, provided the vector of interest is sparse enough ~\cite{bruckstein_elad_zibulevsky2008,slawski_hein_spars2011} (see also~\cite{donoho_tanner2005}).
This finding is key since it implies that compressive PSD estimates can be obtained in a manner similar to noncompressive estimates, i.e., they can be obtained using NNLS (more about this point in a moment).
 
The application of Corollary~\ref{thm:donoho_tanner} to~\eqref{equ:psd_est14} is straightforward.
The required matrix $\B{A}$ is a permuted version of $\widetilde{\B{\Psi}}$ provided the sampling pattern is chosen such that the set of differences $(c_a-c_b)$ form the consecutive sequence $0,1,2,\dotsc,s-1$.
In particular, $\B{A}$ can be constructed from $\B{\Psi}$ as
\begin{equation}
\B{A}_{k,n}=\begin{cases} \Re\{\B{\Psi}_{k/2,n}\} & k=0, 2, 4,\ldots \\
\Im\{-\B{\Psi}_{(k+1)/2,n}\}  & k=1, 3, 5,\ldots 
\end{cases}
\end{equation}
The value for $q$ is chosen such that $q(q-1)+1\geq 2s$, i.e., such that the number of measurements (length of $\uextend$) is greater than twice the sparsity.

One strategy to construct a sampling pattern that forms the difference set $\{0,1,2,\dotsc,s-1\}$ is to again use Golomb rulers. 
Most Golomb rulers do not produce a difference set that is a consequence sequence up to its length, but because differences do not repeat for Golomb rulers, they tend to furnish nearly consecutive sequences.
For example, the order-7 minimum length ruler $\{1, 3, 4, 11, 17, 22, 26\}$ produces the difference set:
\begin{equation}\begin{split}\label{equ:diff_seq}
\{0, 1, &2, 3, 4, 5, 6, 7, 8 , 9, 10, 11,\\
 &13, 14, 15, 16, 18, 19, 21, 22, 23, 25\}
\end{split}
\end{equation} 
which nearly forms the consecutive sequence $0,\ldots,25$, missing only the elements $12, 17, \text{and~} 20$.
In Section~\ref{subsect:ex_sparse_multiband}, we present an example that uses this sampling pattern to successfully recover a 16-sparse vector $\widehat{\B{v}}_C$.
If an appropriate Golomb ruler (sampling pattern) does not exist for a given $L$ and $s$, one can use other CS recovery approaches (e.g., $\ell_1$ minimization, CoSaMP~\cite{needell_tropp2008}, or iterative hard thesholding~\cite{blumensath_davies2009}) with a randomly generated sampling pattern.

The non-negativity requirement of Corollary~\ref{thm:donoho_tanner} for $\widehat{\B{v}}$ is met asymptotically as $N\rightarrow \infty$.
This claim can be inferred from Section~\ref{sect:psd_est} where it is shown that the entries of $\widehat{\B{v}}$ approach integrated periodogram estimates, which are non-negative by construction (see~\eqref{equ:psd_est9} through~\eqref{equ:psd_est11}).
For PSD estimation, the non-negativity requirement is natural since by definition power spectra are non-negative. 

When Corollary~\ref{thm:donoho_tanner} can be applied, we compute compressive estimates $\widehat{\B{v}}_C$ using NNLS:
\begin{equation}\label{equ:psd_est16}
\widehat{\B{v}}_{C}=\arg\min_{\boldsymbol{\alpha}} \norm{\widetilde{\B{\Psi}}\boldsymbol{\alpha}-\uextend}^{2}_{2} \text{~subject to~}\boldsymbol{\alpha}\geq0,
\end{equation}
which is an optimization problem that can be solved via a linear program.
Choosing a NNLS approach for compressive estimates is advantageous because noncompressive estimates can also be computed using a NNLS algorithm (and is perhaps preferred since $\widehat{\B{v}}_{NC}$ can be negative; see Fig.~\ref{fig:sparsePSDex}).

\subsection{Tradeoffs in noncompressive case}\label{subsect:noncomp_tradeoffs}
\begin{figure*}
\begin{minipage}{.48\linewidth}
\centerline{\includegraphics[width=7cm]{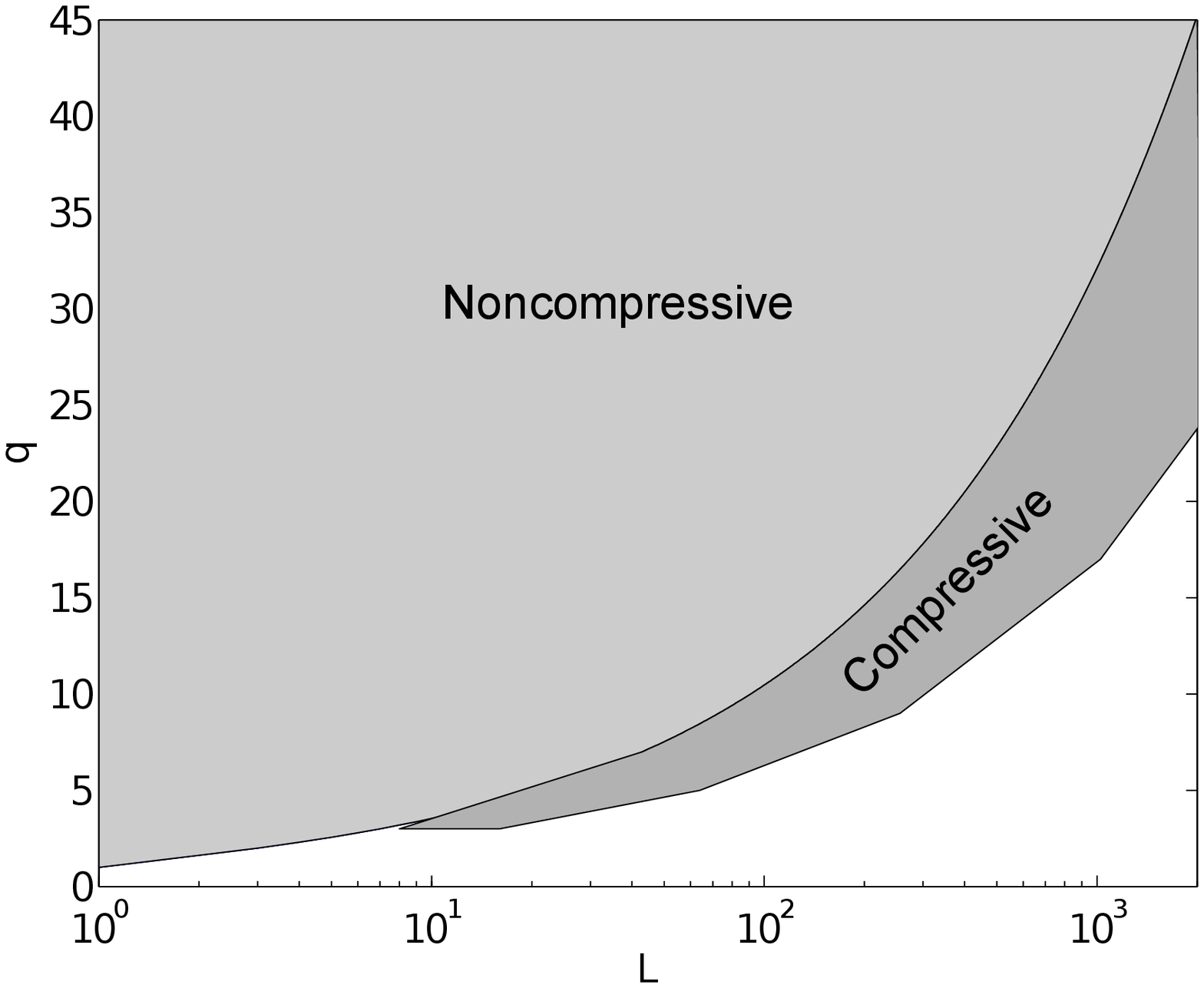}}
\end{minipage}
\begin{minipage}{.48\linewidth}
\centerline{\includegraphics[width=7cm]{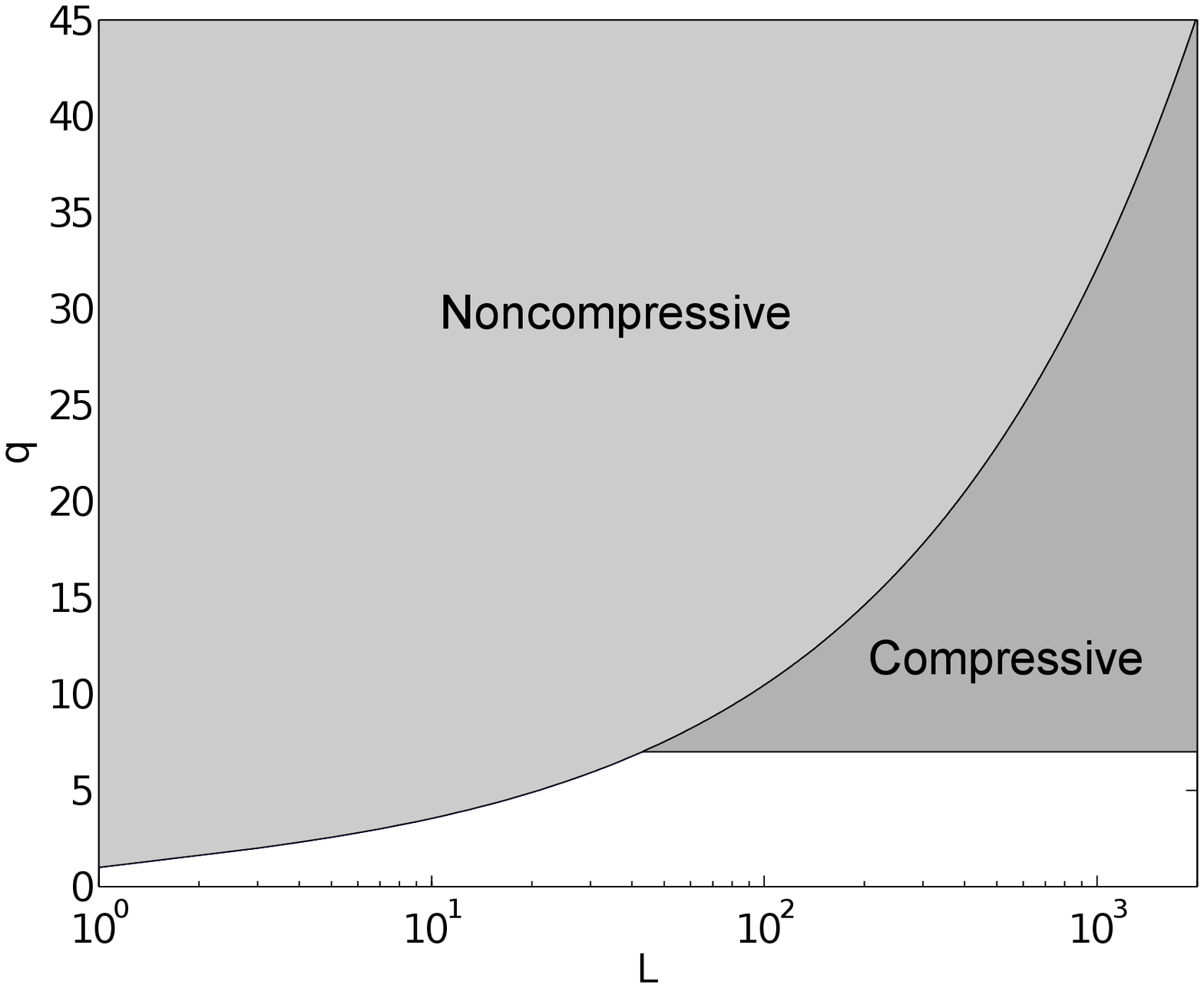}}
\end{minipage}
\caption{Tradeoffs between parameters $q$ and $L$. The lightly shaded areas in both plots demarcate regions in the parameter space where noncompressive estimates exist given that $\widetilde{\B{\Psi}}$ is full rank. The bounding curve is $q(q-1)+1=L$. The darker shaded areas are regions where compressive estimates exist and represent the possible gains in terms of improved parameter tradeoffs over the noncompressive case. The compressive regions derive from the specific scenario where $\widehat{\B{v}}$ is 16-sparse at resolution $L=128$.}\label{fig:qL_tradeoffs1}
\end{figure*}

\begin{figure*}
\begin{minipage}{.48\linewidth}
\centerline{\includegraphics[width=7cm]{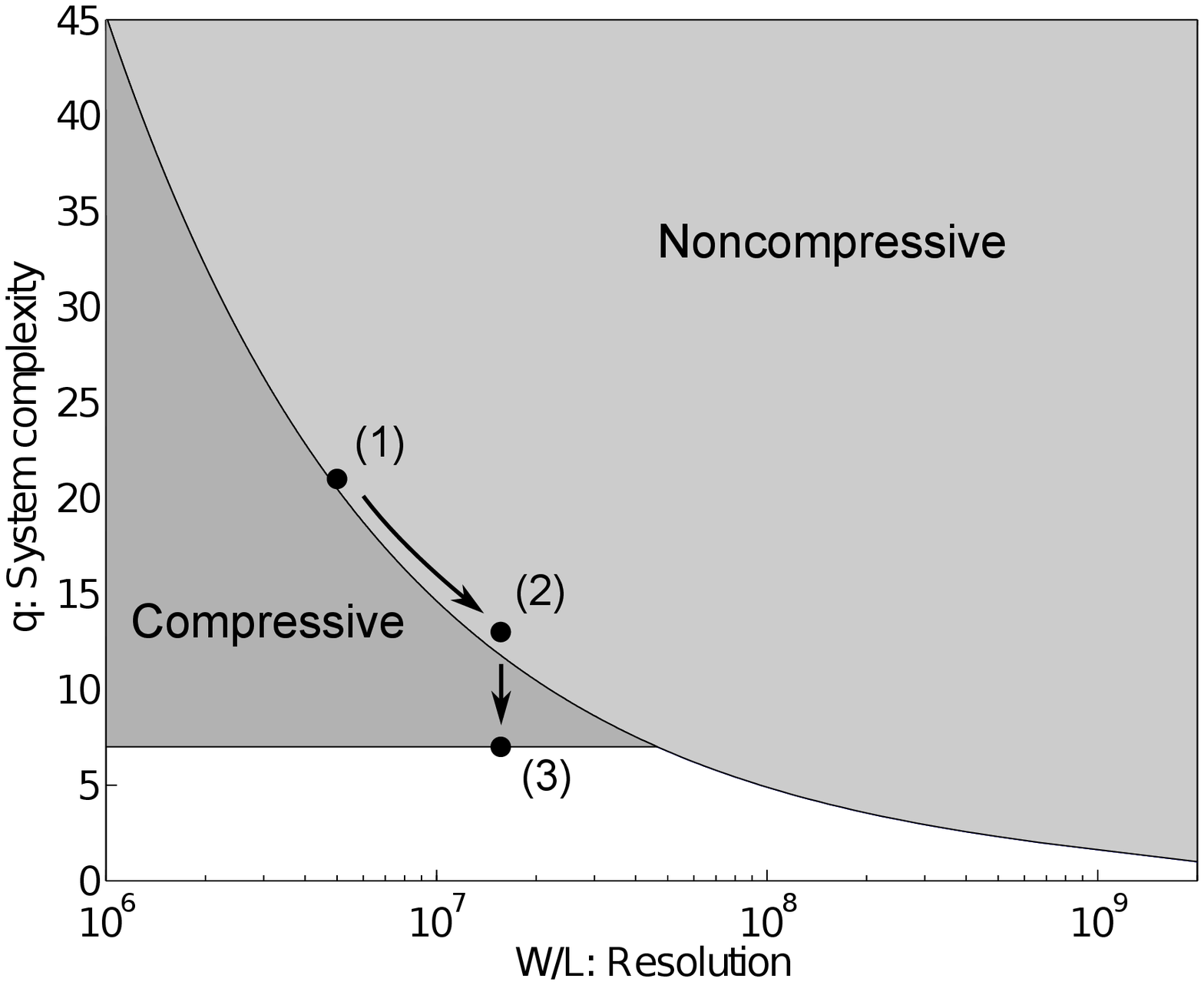}}
\end{minipage}\hfill
\begin{minipage}{.48\linewidth}
\centerline{\includegraphics[width=7cm]{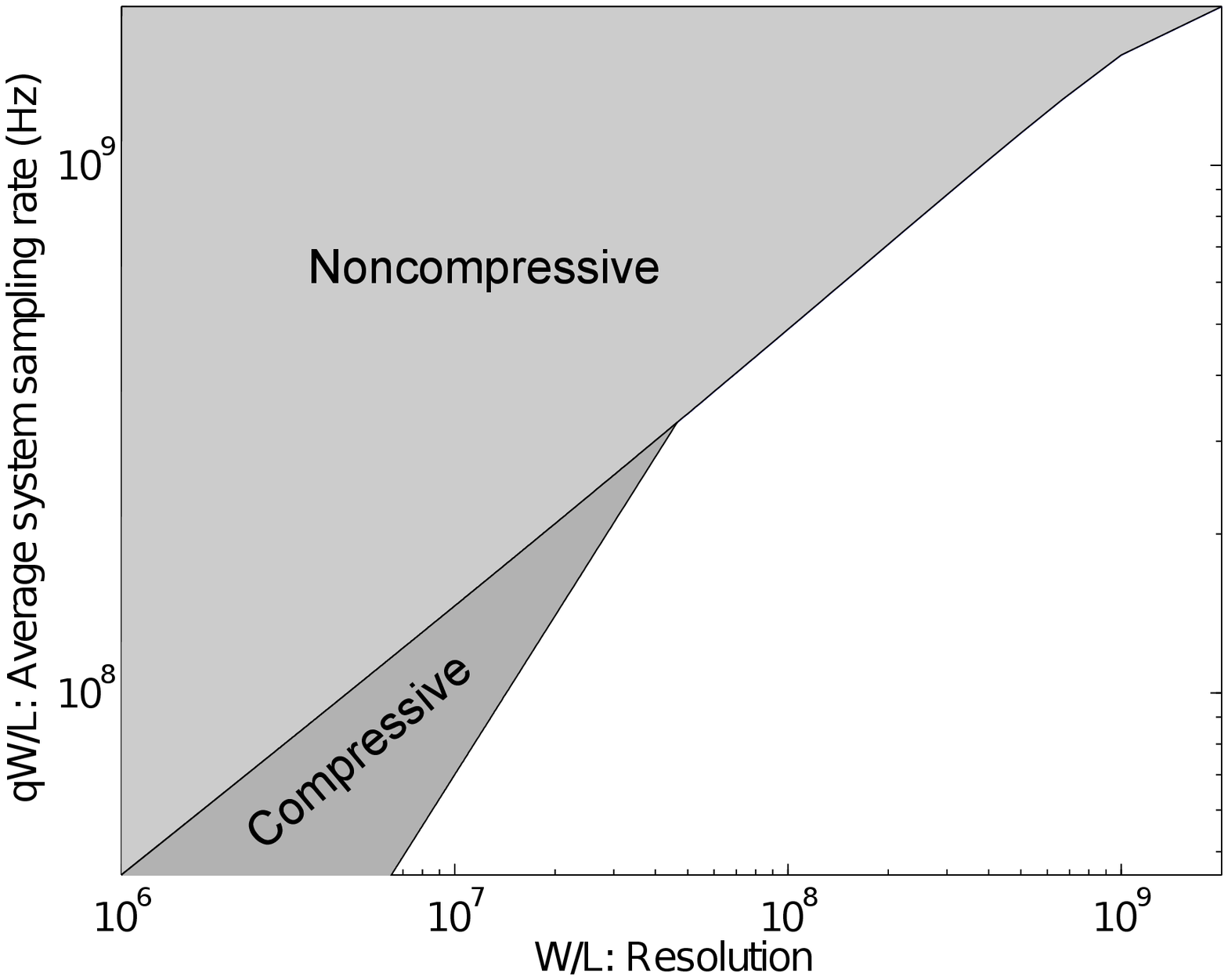}}
\end{minipage}
\caption{Tradeoffs among resolution, system complexity, and system sampling rate. Left: The progression from point (1) to point (2) shows that for a noncompressive estimator the reduction in the number of channels comes at a price of coarser resolution. However, for a 16-sparse vector $\widehat{\B{v}}$ at resolution 15 MHz, a compressive estimator can reduce the number of channels to 7---progression from (2) to (3). Right: Corresponding tradeoff between resolution and system sampling rate.}\label{fig:qL_tradeoffs2}
\end{figure*}

Given a sampling pattern that produces a full column rank $\widetilde{\B{\Psi}}$, the relation $q(q-1)+1\geq L$ establishes a tradeoff between $q$ and $L$ (Fig.~\ref{fig:qL_tradeoffs1}) and tradeoffs among other related quantities (Fig.~\ref{fig:qL_tradeoffs2}).
For example, it establishes a tradeoff between system complexity and the estimate's resolution, where here system complexity is simply taken to be the number of channels $q$.
As the left hand plot in Fig.~\ref{fig:qL_tradeoffs2} shows, finer resolution (smaller $W/L$) comes at the price of higher system complexity.
As a concrete example, consider a WSS signal $x(t)$ band-limited to 1 GHz and a desired resolution of 5 MHz, implying $L=400$.
A noncompressive solution would then require at least 21 channels; however, if a resolution of 15 MHz suffices, the number of channels could be reduced to 13 (see Fig.~\ref{fig:qL_tradeoffs2}).

If one has the the freedom to choose $q$ and $L$, the ratio $q/L$ can be driven to zero even while maintaining the inequality $q(q-1)+1\geq L$.
This means that noncompressive estimates can theoretically be computed at arbitrarily low sampling rates and at finer and finer resolutions (see right hand plot of Fig.~\ref{fig:qL_tradeoffs2}).
Estimation at arbitrarily low sampling rates is a property shared with exisiting \emph{alias-free} PSD estimators~\cite{ahmad_tarczynski2010,tarczynski_qu2005,masry1978a,masry1978b,beutler1970} that are based on random sampling where the set of sampling instances are (typically) realizations of a stochastic point process (e.g., a Poisson point process).
Arbitrarily low sampling rates and arbitrarily fine resolutions, however, require arbitrarily high numbers of channels and arbitrarily long signal acquisition times.
Thus in practical noncompresive situations, an appropriate balance needs to be struck between $q$ and $L$ that satisfies the requirements of a given application.
Section~\ref{subsect:ma_lines} provides evidence of this tradeoff in a specific example. 

\subsection{Improved tradeoffs in compressive case}\label{subsect:comp_tradeoffs}
With sufficient sparsity,~\eqref{equ:psd_est14} can be uniquely solved even if it is underdetermined, i.e., even if ${q (q-1)+1 < L}$~\cite{candes_romberg_tao2006,donoho2006,candes_wakin2008}.
This means that compressive estimates $\widehat{\B{v}}_{C}$ can be computed with better $q-L$ tradeoffs compared to noncompressive ones, i.e., for a given $L$, a compressive estimate can be recovered with a smaller $q$ than what is possible for a noncompressive estimate, or conversely, for a given $q$, a compressive estimate can be recovered with a larger $L$.

The degree of the gain depends on the level of sparsity of $\widehat{\B{v}}$ at a specific resolution $L$.
Let $s$ denote the sparsity (number of nonzero elements) of $\widehat{\B{v}}$ at resolution $L$.
Then according to Corollary~\ref{thm:donoho_tanner}, a compressive estimate can be computed with $2s$ or more Fourier measurements.
Because the number of measurements in~\eqref{equ:psd_est14} equals $q(q-1)+1$ (length of $\uextend$), the need for $2s$ measurements sets a minimum $q$. 
Since compressive estimates pertain to underdetermined linear systems, the condition $q(q-1)+1<L$ sets a maximum $q$.
To apply Corollary~\ref{thm:donoho_tanner}, however, a sampling pattern of length $q$ must exist that produces the set of differences $0,...,s-1$.
If sampling pattern does not exist for the minimum $q$ value, larger $q$ values must be considered.
Thus, for a compressive estimate, the value of $q$ depends on the sparsity of $\widehat{\B{v}}$ (for a given $L$) and on the availability ability of an appropriate sampling pattern.
We discuss specific examples in the next paragraph.
Note that if an alternative CS recovery algorithm is used to solve~\eqref{equ:psd_est14} instead of applying Corollary~\ref{thm:donoho_tanner}, larger $q$ values will be needed for a given $s$ (and $L$) because these algorithms required larger numbers of measurements.  
 
We illustrate the improved tradeoffs in two scenarios each of which are based on the presumption that $\widehat{\B{v}}$ is a 16-sparse vector at resolution  $L=128$ ($\widehat{\B{v}}\in \real^{128}$). 
The results are graphically shown in Fig.~\ref{fig:qL_tradeoffs1}.
As described above, the light shaded regions represent the $(L,q)$ pairs for which noncompressive estimates can be computed%
\footnote{Again, the ability to compute a noncompressive estimate for a valid $(L,q)$ pair also depends on whether a suitable sampling pattern can be found.}%
, and are bordered by the curve $q(q-1)+1=L$.
The darked shaded regions in each panel represent the $(L,q)$ pairs for which compressive estimates can be computed.
The difference between the compressive regions on the left and right plots is that the region on the left presumes $s$ changes with $L$ and the one on the right presumes $s$ is independent of $L$, i.e., it remains constant as $L$ varies.
In the left hand plot, we assume $s$ doubles every time $L$ doubles.
Thus, in comparison to the initial presumption $s=16$ at $L=128$,  we assume $s=32$ at $L=256$ which implies a mimimum $q=9$.
This behavior models the situation where the active subbands of $x(t)$ at resolution $L=128$ contain energy at every frequency within the subband, so as the subbands are subdivided, $s$ increases.
The scenario on the right reflects the situation where the active subbands only contain energy at one specific frequency, so $s$ remains constant as $L$ increases.

To judge the implications of these tradeoffs, we examine the associated tradeoffs among system complexity ($q$), resolution ($W/L$), and average system sampling rate ($qW/L$) in Fig.~\ref{fig:qL_tradeoffs2}.
As above, the compressive regions are drawn under the presumption that $s=16$ for $L=128$.
The plots, however, are only for the case where $s$ is independent of $L$.
Intuitively, this represents a best case scenario in terms of ``CS gain'' over the noncompressive case (best possible improvements in $q$ for a fixed $L$ and in $L$ for a fixed $q$).
The left hand panel  shows system complexity as a function of resolution $W/L$ where $W=2$ GHz. 
The right hand panel plots resolution versus system sampling rate.
The plots show that for this particular scenario, a CS approach can have appreciable benefits when the underlying PSD is sparse.
For example, in Section~\ref{subsect:noncomp_tradeoffs} we stated that the recovery of $\widehat{\B{v}}$ with a resolution of 15 MHz required a 13 channel MC sampler for a noncompressive estimator.
However, for a 16-sparse vector $\widehat{\B{v}}$, a compressive estimate with the same resolution can be recovered with only $q=7$ channels, a reduction of almost 1/2.
This gain is depicted in Fig.~\ref{fig:qL_tradeoffs2} as the shift from point (2) to point (3).
The same reduction factor can also be seen in average system sampling rate (right hand plot of Fig.~\ref{fig:qL_tradeoffs2}).
 
These improved tradeoffs may or may not effect the error in the recovered estimate.
If $L$ is held fixed and $\widehat{\B{v}}$ is sparse, then as described above, $q$ can be reduced to the smallest integer such that $q(q-1)+1>2s$ (according to Corollary~\ref{thm:donoho_tanner}).
This reduction in $q$ reduces the overall system sampling rate; however, the error in the recovered compressive estimate is \emph{not} significantly different from the error in a corresponding noncompressive estimate because CS theory guarantees accurate (if not exact) recovery of $\widehat{\B{v}}$.
Now, if $q$ is held fixed and $L$ is increased, the sampling rate of each channel decreases.
Thus, if the observation interval is fixed, the correlation estimates $\widehat{u}_i$ become worse (in the sense that their error is larger) because less and less samples are used to compute $\widehat{u}_i$.
In this case, CS still allows the recovery of an estimate, but its quality is declining as $L$ increases.
To maintain accuracy, one must increase the observation interval and collect more samples per channel.

\subsection{Consistency of the noncompressive estimator}\label{subsect:bias_var}
To simplify notation, we examine the bias and variance of the least squares solution to $\widehat{\B{u}}=\B{\Psi} \widehat{\B{v}}$ instead of the solution of the expanded version $\uextend= \widetilde{\B{\Psi}} \widehat{\B{v}}$.
The analysis is equivalent in either case.
Assuming $\B{\Psi}$ is full rank, the noncompressive solution is 
\begin{align*}
\widehat{\B{v}}_{NC} &= (\B{\Psi}^{H}\B{\Psi})^{-1}\B{\Psi}^{H}\widehat{\B{u}} \\
&= \B{\Psi}^{\dagger}\widehat{\B{u}},
\end{align*}
where $H$ denotes Hermitian transpose and $\B{\Psi}^{\dagger}$ is shorthand notation for the pseudoinverse.

\textbf{Bias.} The bias of $\widehat{\B{v}}_{NC}$ is the amount its expected value differs from $\B{v}_{NC}=\B{\Psi}^{\dagger}\B{u}$.
Taking the expectation, we have $\expect \widehat{\B{v}}_{NC}=\B{\Psi}^{\dagger}\expect\widehat{\B{u}}$.
One element of $\expect \widehat{\B{u}}$ equals
\begin{align}\label{equ:bias0}
\expect \widehat{u}_i &=\,\expect \widehat{r}_{z_az_b}^{\,N}(0) \\
\begin{split}
&=\frac{1}{N} \sum_{m=-\infty}^{\infty}\sum_{l=-\infty}^{\infty}h_a(m) h_b(l) \\ 
&\quad  \sum_{n=0}^{N-1} \expect[y_a(n-m)y_b(n-l)] w_R(n-m)w_R(n-l) \end{split} \\
\begin{split}
&=\sum_{m}\sum_{l}h_a(m) h_b(l) r_{y_ay_b}(l-m) \\
&\qquad \frac{1}{N} \sum_{n=0}^{N-1} w_R(n-m)w_R(n-l). \end{split}\label{equ:noncomp_bias}
\end{align}
where, as in~\eqref{equ:psd_est13}, $i$ is an index for the pair $(a,b)$, the second equality follows from~\eqref{equ:psd_est2} and~\eqref{equ:psd_est3}, and $w_R(n)$ is a discrete rectangular window
\begin{equation*}
w_{R}(n)=\begin{cases}1 \qquad 0\leq n \leq N-1 \\
0 \qquad \text{otherwise}.  \end{cases}
\end{equation*}
It is straightforward to show that
\begin{equation}\begin{split}
\frac{1}{N} \sum_{n=0}^{N-1} &w_R(n-m)w_R(n-l) \\
&=\begin{cases} 1-\frac{\lvert m-l \rvert}{N} &\lvert m-l \rvert \leq N-1 \\
0 &\lvert m-l \rvert > N-1. \end{cases}\end{split}
\end{equation}
By substituting this expression back into~\eqref{equ:noncomp_bias}, it is clear that $\widehat{\B{v}}_{NC}$ is a biased estimate for finite $N$; however asympotically, $\widehat{\B{v}}_{NC}$ is unbiased since
\begin{align}
\expect \widehat{u}_i &\overset{N\rightarrow \infty}{=}\sum_{m}\sum_{l}h_a(m) h_b(l) r_{y_ay_b}(l-m)\\
&~~\,=r_{z_az_b}(0) \qquad \text{(from~\eqref{equ:biasref})}\\
&~~\,=u_i. 
\end{align}
implies $\expect \widehat{\B{v}}_{NC} =\B{\Psi}^{\dagger}\expect \widehat{\B{u}} \rightarrow \B{\Psi}^{\dagger}\B{u}=\B{v}$ as $N\rightarrow \infty$.

\textbf{Variance.}
The covariance matrix of $\widehat{\B{v}}_{NC}$ is
\begin{equation}\begin{split}\label{equ:covar_v}
\expect \bigl(\widehat{\B{v}}_{NC}-\expect[\widehat{\B{v}}_{NC}]\bigr)\bigl(\widehat{\B{v}}_{NC}-\expect[\widehat{\B{v}}_{NC}]\bigr)^{H} \\
\quad =\B{\Psi}^{\dagger} \underset{\displaystyle\triangleq\B{K}}{\underbrace{\expect (\widehat{\B{u}}-\expect\widehat{\B{u}})(\widehat{\B{u}}-\expect\widehat{\B{u}})^{H}}}[\B{\Psi}^{\dagger}]^{H}\end{split}
\end{equation}
where $\B{K}$ is the covariance matrix of $\widehat{\B{u}}$.
The diagonal elements of~\eqref{equ:covar_v} are the variances of the elements of $\widehat{\B{v}}_{NC}$:
\begin{equation}\label{equ:covar_v1}
\var{\widehat{v}_{l}}=\sum_{i} \sum_{j} \B{\Psi}^{\dagger}_{l,j}\B{K}_{i,j}\B{\Psi}^{\dagger}_{l,i}, \quad l=0,\dotsc,L-1
\end{equation}
where $\B{K}_{i,j}=\covar{\widehat{u}_{i},\widehat{u}_{j}}$.
Note that in~\eqref{equ:covar_v1} we have dropped the subscript $NC$ in our notation for the elements of $\widehat{\B{v}}_{NC}$.
To show that $\var{\widehat{v}_{l}}\rightarrow 0$ as $N\rightarrow \infty$, we proceed to upper bound~\eqref{equ:covar_v1} with a sequence in $N$ that approaches zero as $N\rightarrow \infty$. 

First, we note that the inequality $[\covar{\widehat{u}_{i},\widehat{u}_{j}}]^2 \leq \var{\widehat{u}_{i}} \var{\widehat{u}_{j}}$~\cite[p. 336]{castella_berger2002} implies
\begin{equation}\label{equ:covar_v2}
\lvert \covar{\widehat{u}_{i},\widehat{u}_{j}}\rvert \leq \sqrt{\var{\widehat{u}_{i}} \var{\widehat{u}_{j}}}.
\end{equation}
Second, for the correlation estimate $\widehat{u}_{i}=\widehat{r}_{z_az_b}^{\,N}(0)$ it is well-known~\cite[p. 320]{johnson_dudgeon1993} that%
\footnote{The reason for the approximation in~\eqref{equ:covar_v3} is that this expression is derived assuming that $z_a(n)$ and $z_b(n)$ are jointly Gaussian; however, the result applies to other distributions as well~\cite{johnson_dudgeon1993}.}
\begin{equation}\label{equ:covar_v3}
\var{\widehat{r}_{z_az_b}^{\,N}(0)}\approx \frac{2}{N} \sum_{m=-(N-1)}^{N-1} \Big(1-\frac{\lvert m \rvert}{N}\Big) r_{z_az_b}^{2}(m).
\end{equation}
Hence, if $r_{z_az_b}^{2}(m)$ is square-summable, $\var{\widehat{r}_{z_az_b}^{\,N}(0)}=\var{\widehat{u}_{i}}$ approaches zero as $N\rightarrow \infty$.
Bound~\eqref{equ:covar_v1} by
\begin{equation}
\var{\widehat{v}_{l}}\leq \sum_{i} \sum_{j} \lvert \B{\Psi}^{\dagger}_{l,j} \rvert ~ \lvert \B{K}_{i,j} \rvert ~ \lvert\B{\Psi}^{\dagger}_{l,i}\rvert.
\end{equation}
Then from~\eqref{equ:covar_v2} and~\eqref{equ:covar_v3}, we conclude that $\var{\widehat{v}_{l}}$, and thus $\var{\widehat{\B{v}}_{NC}}$, approach zero as $N\rightarrow \infty$.   

The noncompressive estimator $\widehat{\B{v}}_{NC}$ is thus statistically consistent since it is asymptotically unbiased and since its variance decreases to zero as $N\rightarrow \infty$.
Incidentally, if $x(t)$ is a zero-mean Gaussian WSS random process, $\widehat{\B{v}}_{NC}$ is also asymptotically efficient~\cite[p. 471]{castella_berger2002}, meaning that the variance of the limit distribution of $\widehat{\B{v}}_{NC}$ achieves the Cram\'{e}r-Rao bound.
As we show in the Appendix, asymptotic efficiency follows from the fact that $\widehat{\B{u}}$ is a maximum likelihood estimate of $\B{u}$.

\subsection{Consistency of the compressive estimator}\label{subsect:comp_consistency}
A detailed analysis of the compressive estimator's consistency is beyond the scope of this paper because it depends on the specific CS recovery algorithm is employed.
Nevertheless, compressive estimates are consistent in the sense that ideally sparse vectors $\widehat{\B{v}}$ can be \emph{exactly} recovered by some CS recovery algorithms, and thus the consistency of $\widehat{\B{u}}$  (as shown above) implies the consistency of $\widehat{\B{v}}_{C}$.

\section{Examples}\label{sect:ex}
\subsection{Estimating MA spectra with lines}\label{subsect:ma_lines}
\begin{figure*}
\begin{minipage}{.98\linewidth}
\subfigure[True power spectrum]{\centerline{\includegraphics[width=7cm]{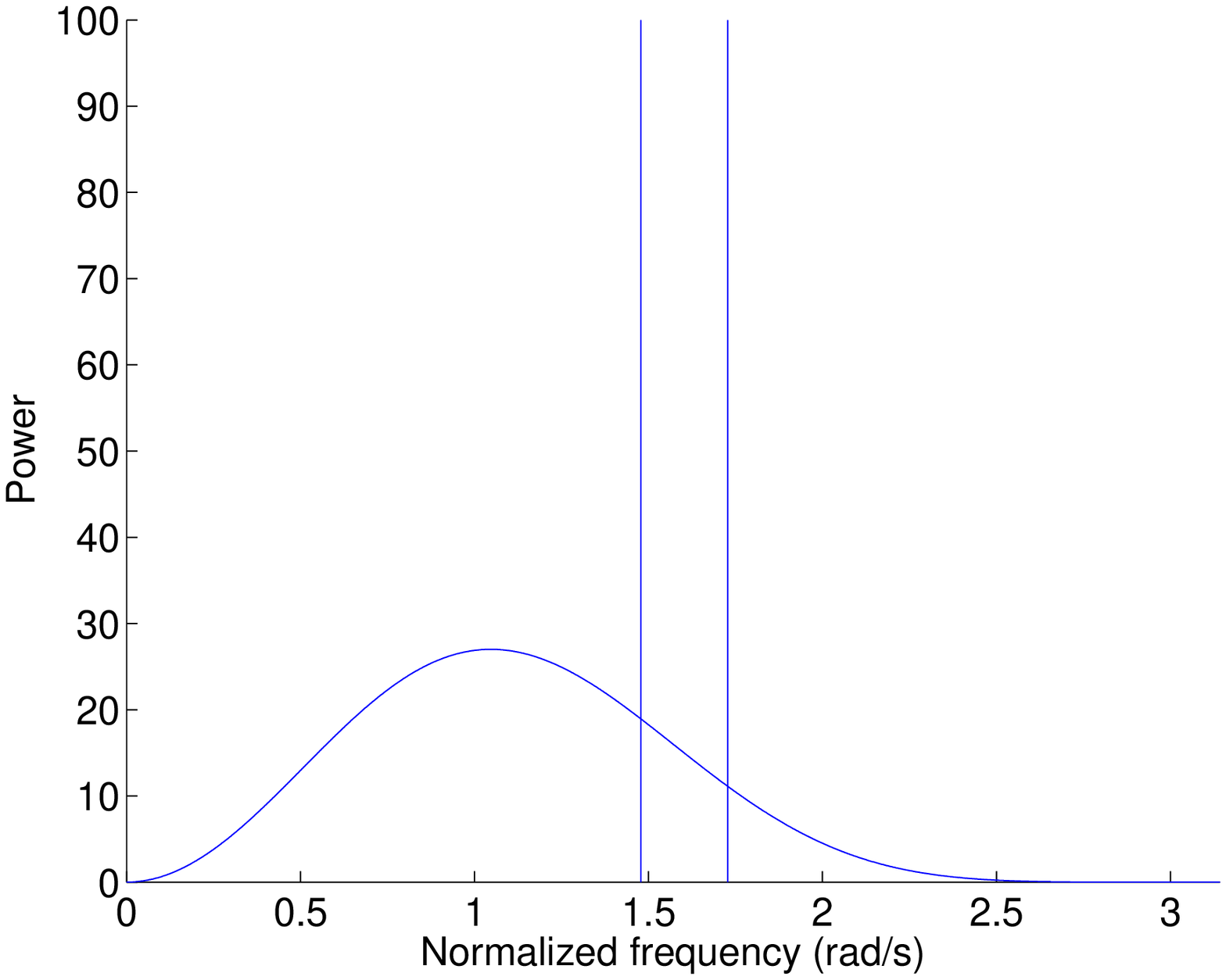}}\label{subfig:ma1}} 
\end{minipage}
\begin{minipage}{0.48\linewidth}
\subfigure[$q\negthinspace=\negthinspace50, L\negthinspace=\negthinspace64, N\negthinspace=\negthinspace50$]{\centerline{\includegraphics[width=7cm]{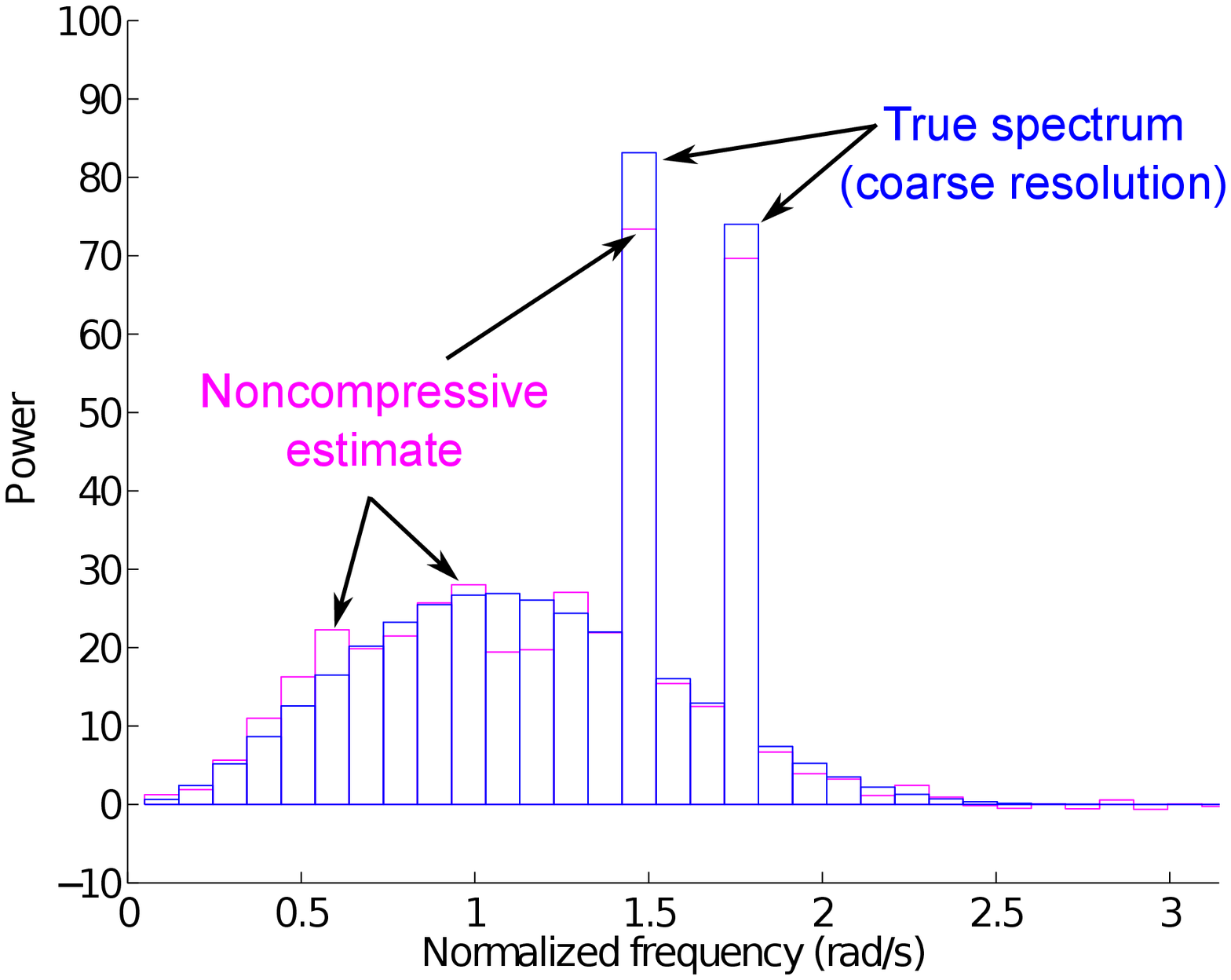}}\label{subfig:ma2}} 
\end{minipage}
\begin{minipage}{.48\linewidth}
\subfigure[$q\negthinspace=\negthinspace50, L\negthinspace=\negthinspace64, N\negthinspace=\negthinspace10000$]{\centerline{\includegraphics[width=7cm]{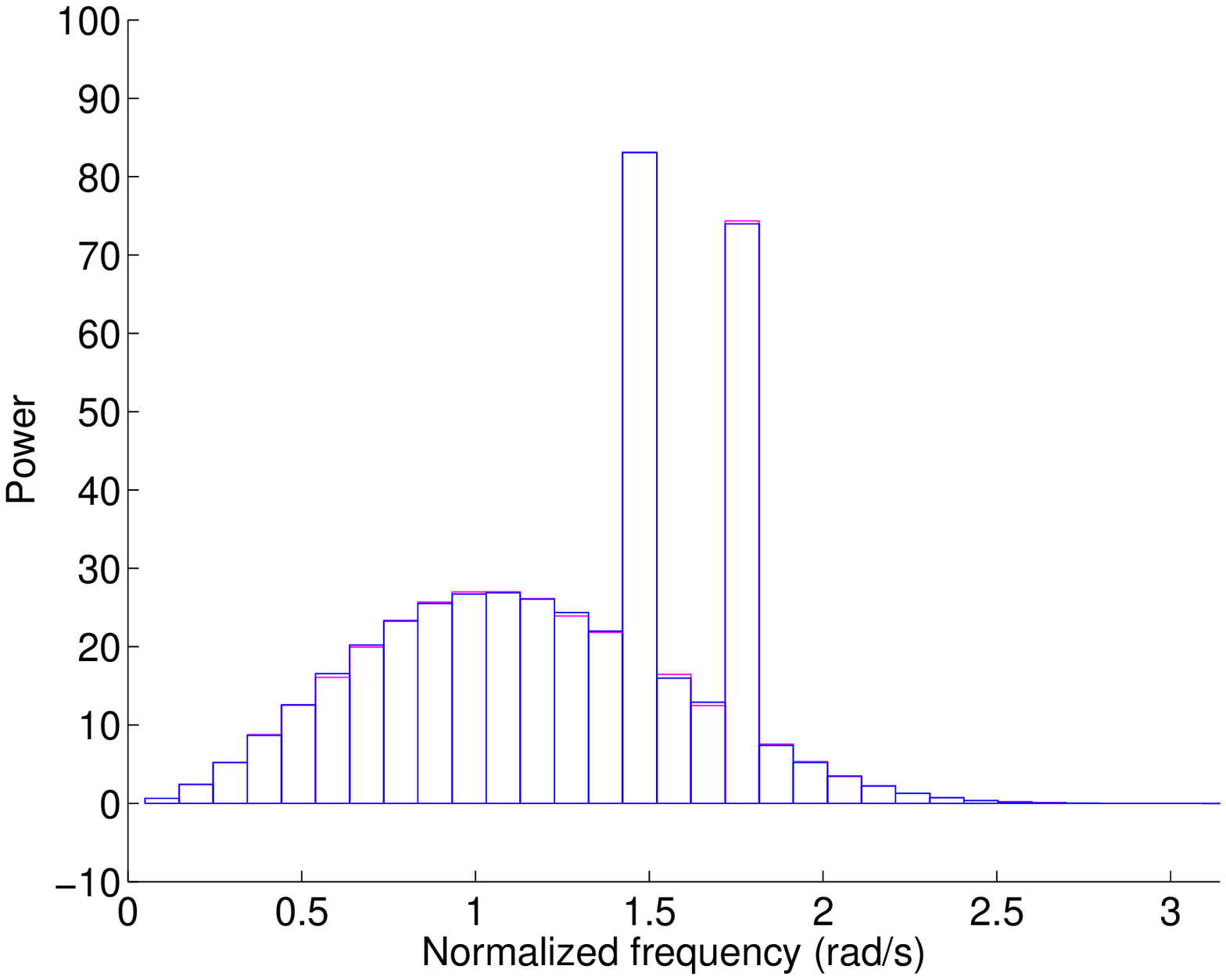}}\label{subfig:ma3}}
\end{minipage}
\caption{Estimating MA spectra with lines. (a) True power spectrum. (b) Overlay plot of a noncompressive estimate and a coarse resolution approximation of the true power spectrum. The height of the bars represent the average power in the band which it spans. (c) Estimate of same spectrum except the number of samples per channel, $N$, increased from 50 to 10000. Note the convergence of the noncompressive estimate.}\label{fig:MA_lines_ex}
\end{figure*}

\begin{figure*}
\begin{minipage}{0.48\linewidth}
\centerline{\includegraphics[width=7cm]{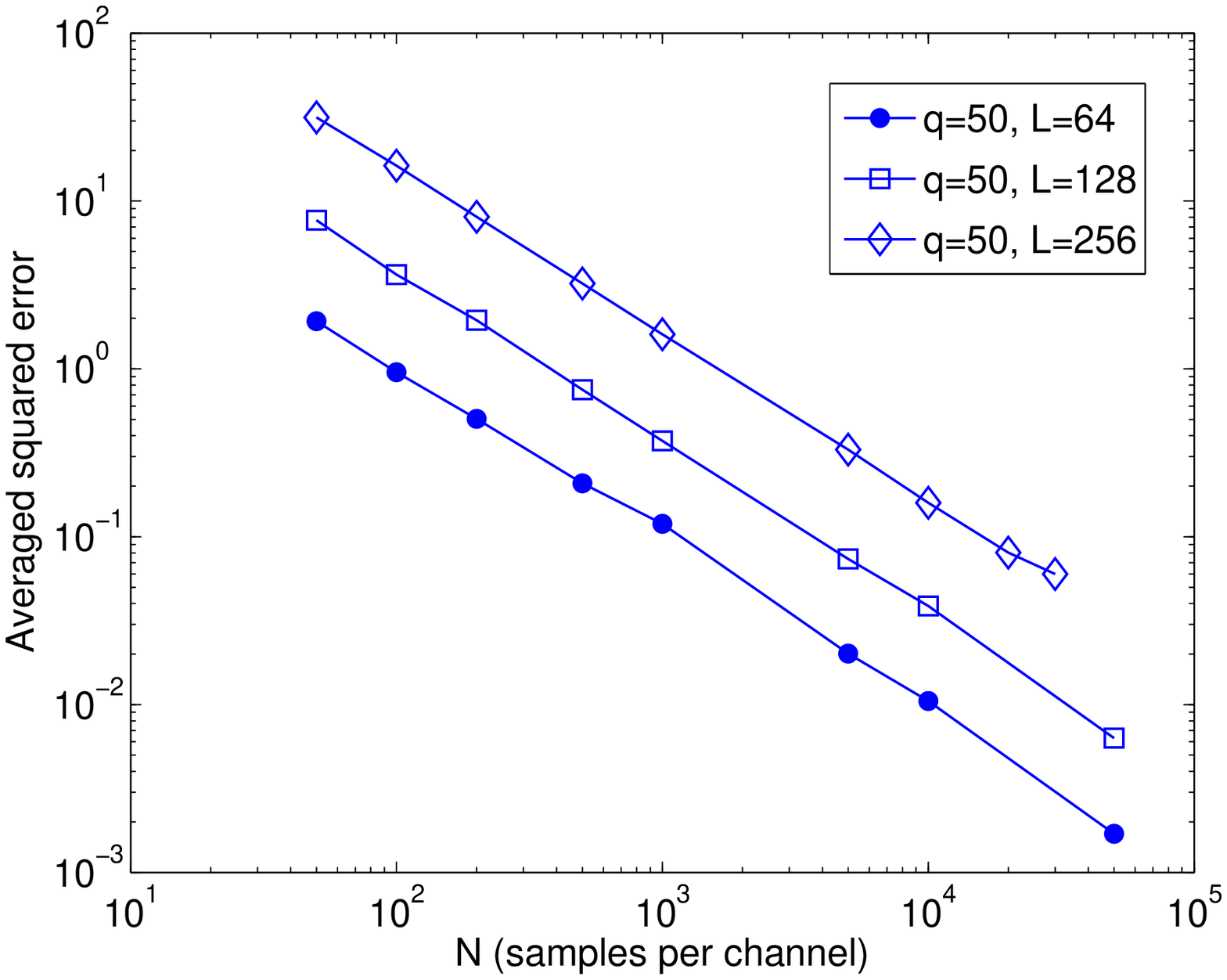}}  
\end{minipage}
\begin{minipage}{0.48\linewidth}
\centerline{\includegraphics[width=7cm]{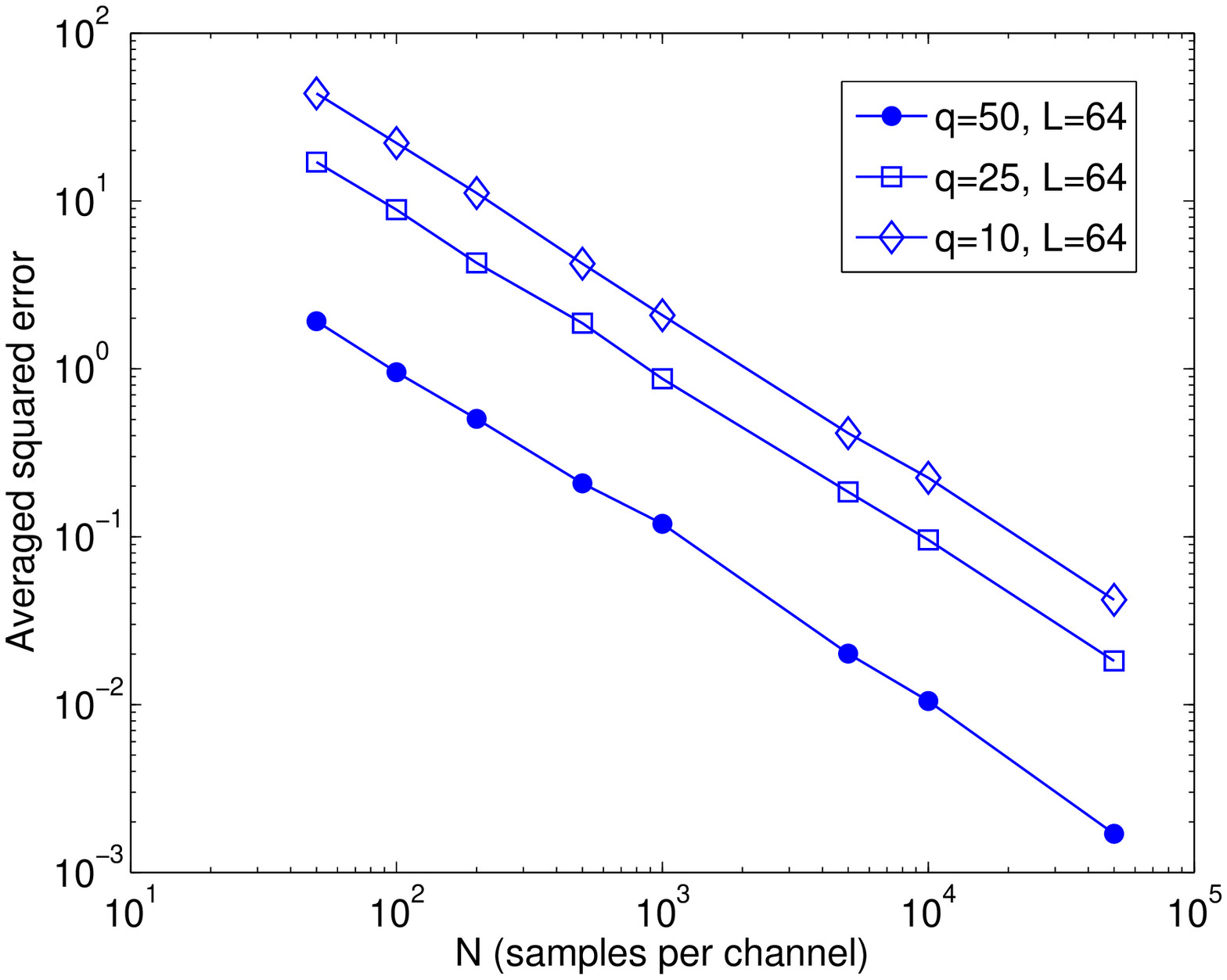}}  
\end{minipage}
\caption{Decrease in average squared error for various noncompressive scenarios as $N\rightarrow \infty$.}\label{fig:maMSE}
\end{figure*}

Consider sampling a (nonsparse) moving average power spectrum with spectral lines using a 50 channel MC sampler ($q=50, L=64$).
The process is generated by passing white noise (with unit variance) through a filter with transfer function $H(z)=(1-z^{-1})(1+z^{-1})^{3}$ and then adding the two sinusoidal components, $2\cos{\bigl(\tfrac{8\pi k}{17}\bigr)}$ and  $2\cos{\bigl(\tfrac{11\pi k}{20}\bigr)}$. 
Fig.~\ref{subfig:ma1} shows the true PSD.
Fig.~\ref{subfig:ma2} and Fig.~\ref{subfig:ma3} show two resulting noncompressive estimates for different values of $N$, the number of MC samples per channel.
For comparison, a coarse resolution approximation of the true PSD is overlaid on the estimates.
Because $\widehat{\B{v}}_{NC}$ is consistent, its mean squared error with respect to this coarse resolution approximation decreases to zero as $N\rightarrow \infty$.
This is depicted in Fig.~\ref{subfig:ma3}.
Fig.~\ref{fig:maMSE} further illustrates the consistency of $\widehat{\B{v}}_{NC}$ within the context of this example for various $(L,q)$ values. 
In each case, the average squared error monotonically decreases as $N$ increases.

\subsection{Sparse multiband power spectra}\label{subsect:ex_sparse_multiband}
Suppose a real WSS random process $x(t)$ has a sparse multiband power spectrum band-limited to $1$ GHz ($W=2$ GHz).
Suppose also that we are interested in a PSD estimate having a resolution of about 15 MHz.
To satisfy this requirement we choose $L=128$ which gives a resolution of 15.625 MHz. 
Suppose also that the sparse PSD is composed of two active bands, each with an approximate bandwidth of 30 MHz.
Being conservative, we thus expect $\B{v}$ to be a 16-sparse vector (including positive and negative frequencies with each active band taking up 4 adjacent subbands).
\begin{figure*}
\begin{minipage}{.48\linewidth}
\subfigure[$q\negthinspace=\negthinspace20, L\negthinspace=\negthinspace128, N\negthinspace=\negthinspace1000$ (overdetermined)]{\centerline{\includegraphics[width=7cm]{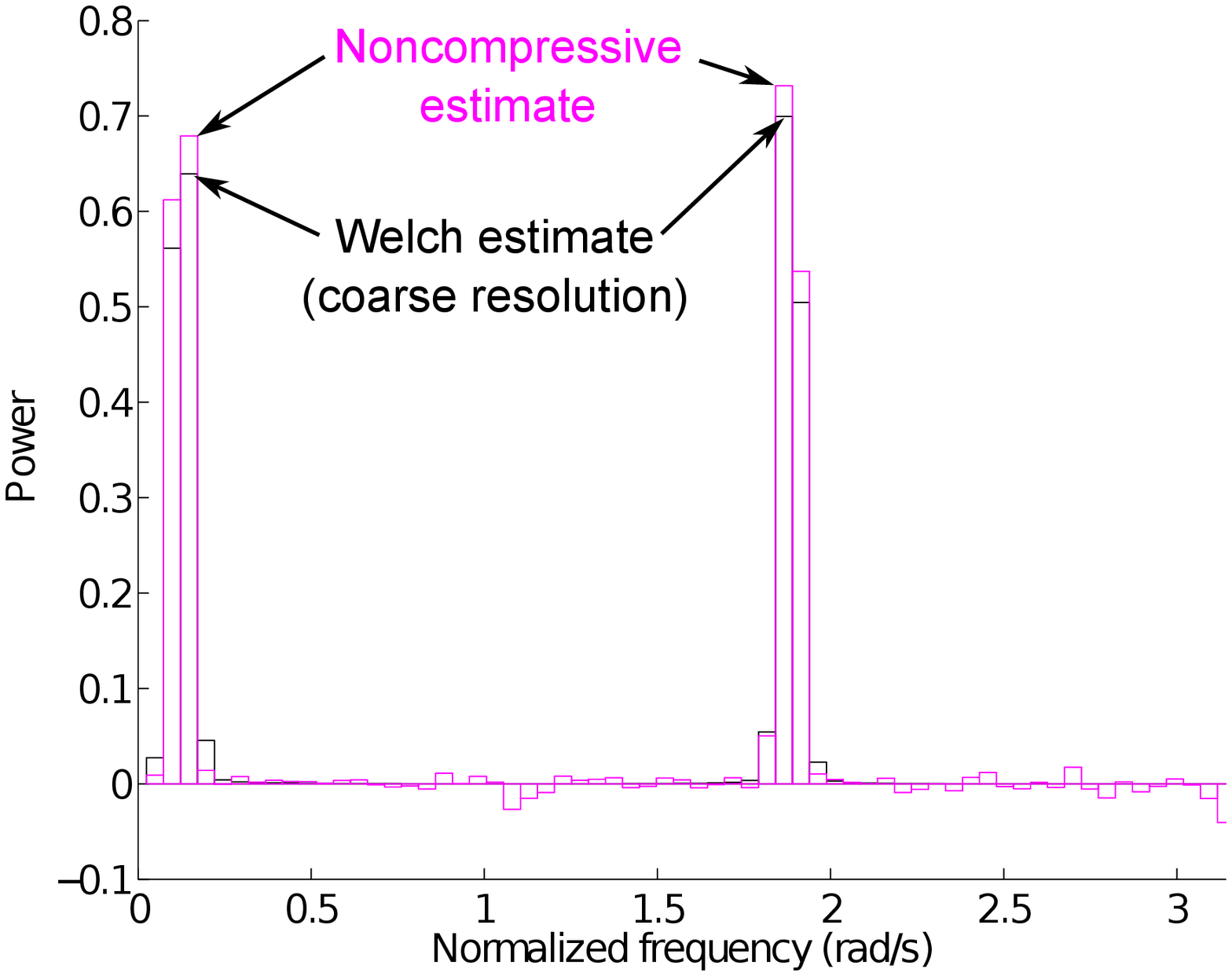}}\label{subfig:ex1a}}
\end{minipage}
\begin{minipage}{.48\linewidth}
\subfigure[$q\negthinspace=\negthinspace7, L=128, N=1000$ (underdetermined)]{\centerline{\includegraphics[width=7cm]{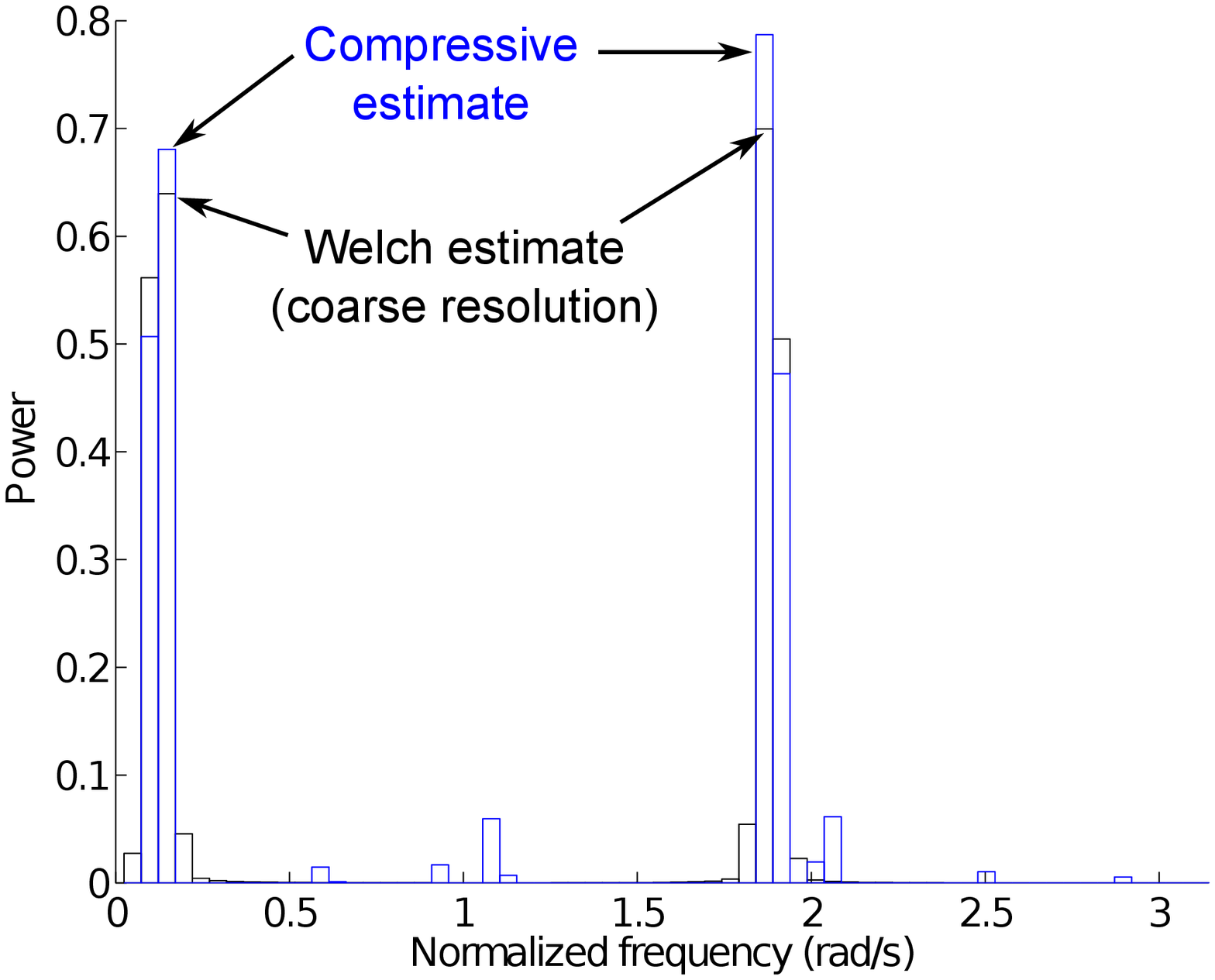}}\label{subfig:ex1b}}
\end{minipage}
\begin{minipage}{.48\linewidth}
\subfigure[$q\negthinspace=\negthinspace7, L\negthinspace=\negthinspace128, N\negthinspace=\negthinspace10000$ (underdetermined)]{\centerline{\includegraphics[width=7cm]{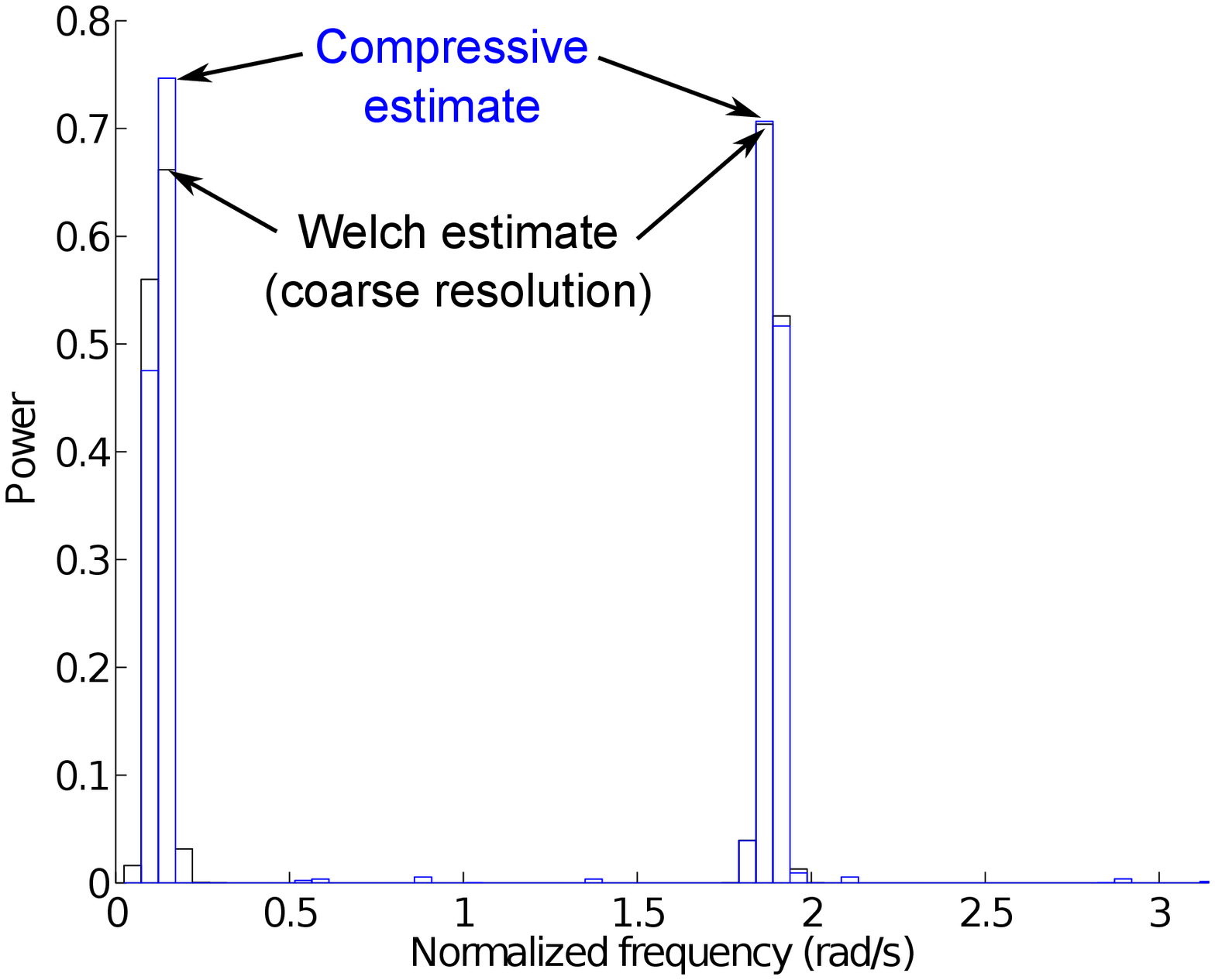}}\label{subfig:ex1c}}
\end{minipage}
\begin{minipage}{.48\linewidth}
\vspace*{3pt}
\subfigure[$q\negthinspace=\negthinspace7, L\negthinspace=\negthinspace128, N\negthinspace=\negthinspace10000$ (underdetermined)]{\centerline{\includegraphics[width=7cm]{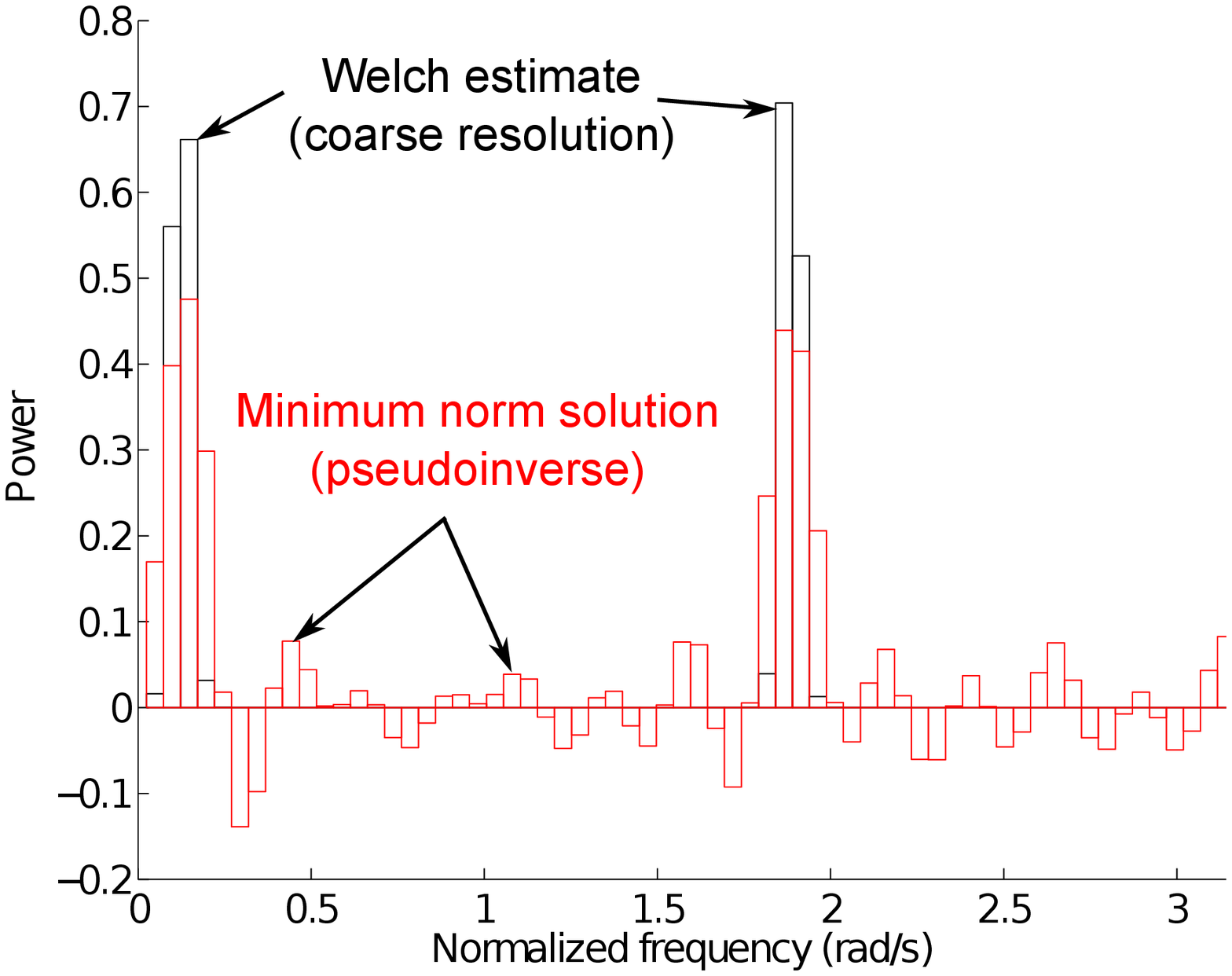}}\label{subfig:ex1d}}
\end{minipage}
\caption{Estimating sparse multiband power spectra. The overlay plots compare compressive and noncompressive estimates to a coarse resolution Welch estimate (black) based on Nyquist rate samples. (a) Noncompressive estimate (magenta) with random sampling pattern. (b)-(c) Compressive estimates (blue) with Golomb ruler sampling pattern. (d) Minimum norm least squares estimate (red) in underdetermined case.}\label{fig:sparsePSDex}
\end{figure*}

The PSD can be estimated in either a noncompressive or a compressive manner.
Choosing $q=20$, leads to an overdetermined system in~\eqref{equ:psd_est7} ($q(q-1)+1=381 \geq 128=L$) and a noncompressive estimate.
Fig.~\ref{subfig:ex1a} shows the resulting noncompressive estimate with a normalized square error of $0.010$.
The error is calculated with respect to a coarse Welch estimate, i.e., it has the same resolution as the noncompressive estimate.
The Welch estimate is based on uniform Nyquist rate samples and is computed with no overlap.
The sampling pattern was chosen uniformly at random from the set $\{0,\ldots,127\}$ and it was verified that $\widetilde{\B{\Psi}}$ had full rank.
The fractional delays were implemented by interpolating the MC sequences to the Nyquist rate, shifting them by the appropriate delays, and then subsampling them to the original rate.
Each channel collected 1000 MC samples and the average system sampling rate was about 1/6 of the Nyquist rate.

According to Corollary~\ref{thm:donoho_tanner}, the number of measurements required to recover a 16-sparse vector needs to be greater than or equal to 32.
This implies that $q$ needs to be at least 7 because the number of measurements in~\eqref{equ:psd_est14} equals $q(q-1)+1$ and for $q<7$ this quantity is less than 32.
In addition, Corollary~\ref{thm:donoho_tanner} requires that the sampling pattern produces the set of differences $0,\ldots,15$ for a 16-sparse signal.
Fortunately, a Golomb ruler of order 7 ($\{c_i\}=\{1, 3, 4, 11, 17, 22, 26\}$) exists that produces the necessary difference set.
Fig.~\ref{subfig:ex1b} shows the resultant compressive estimate (normalized squared error $0.023$).
The reduction in the number of channels from 13 to 7, in going from the noncompressive to the compressive estimator, reduces the complexity of the sampler, or equivalently, halves the average system sampling rate.

Fig.~\ref{subfig:ex1c} displays the same compressive estimate as in Fig.~\ref{subfig:ex1b} except that it is computed with 10000 samples per channel instead of 1000.
The normalized squared error is $0.013$.
In contrast to this figure, Fig.~\ref{subfig:ex1d} shows the minimum ($\ell_2$) norm least squares solution computed by the pseudoinverse.
As one would expect, this solution is far from the true PSD and does not improve with more samples.

\subsection{Spectral sensing for cognitive radio}
To opportunistically use their resources, cognitive radios must be able to dynamically sense underutilized portions of the radio spectrum.
Towards this end, methods and algorithms have recently been proposed to sense the largest possible bandwidth while sampling at the lowest possible rate~\cite{ariananda_leus2011}.
Some of these methods have taken advantage of CS, but the following example shows that a noncompressive estimator can monitor large bandwidths at low sampling rates (although, as explained above, CS will allow better tradeoffs with a compressive estimator).

\begin{figure*}
\begin{minipage}{0.48\linewidth}
\subfigure[$q\negthinspace=\negthinspace25, L\negthinspace=\negthinspace64, N\negthinspace=\negthinspace4096$]{\centerline{\includegraphics[width=7cm]{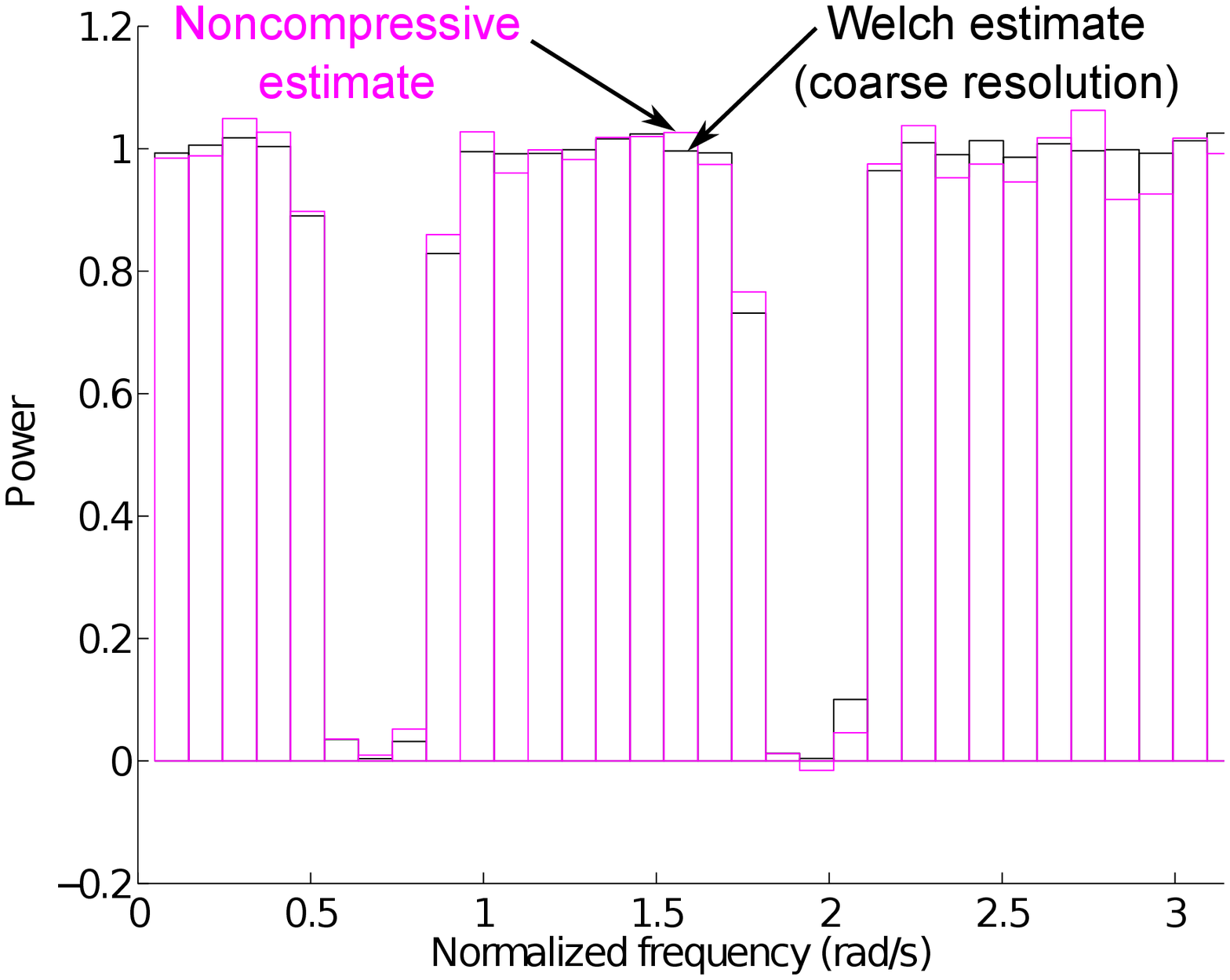}}}
\end{minipage}
\begin{minipage}{0.48\linewidth}
\subfigure[$q\negthinspace=\negthinspace50, L\negthinspace=\negthinspace128, N\negthinspace=\negthinspace2048$]{\centerline{\includegraphics[width=7cm]{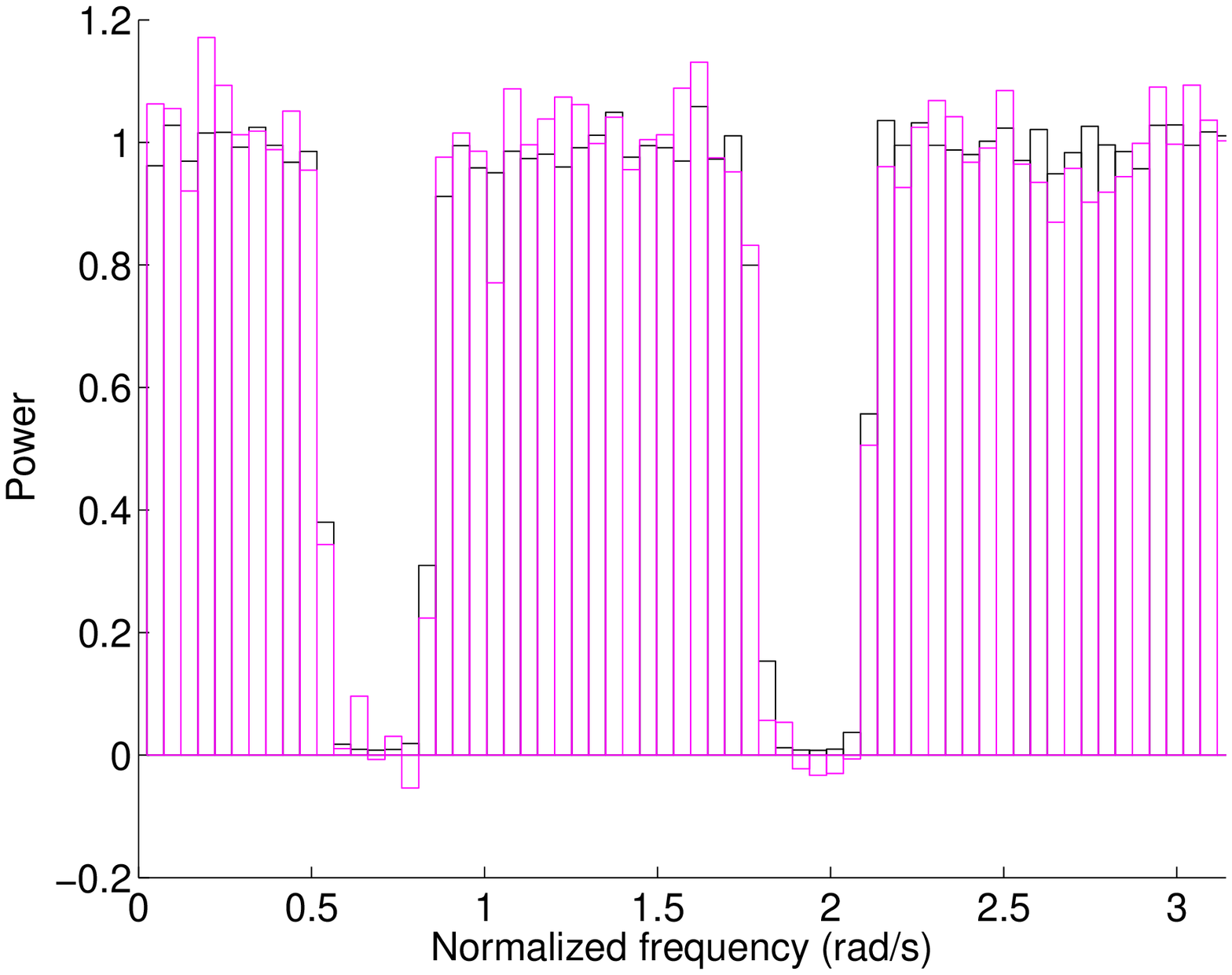}}}
\end{minipage}
\caption{Estimating spectra with holes. The plots show two different noncompressive estimates (differring resolutions) of a MA spectrum with two spectral notches (holes).}\label{fig:cognitive_radio_ex}
\end{figure*}

This example is only a caricature of the actual problem that must be solved to realize a practical cognitive radio system.
Let $x(t)$ be a MA random process generated from filtering white Gaussian noise with a notch filter.
The filter notches the spectrum such that it has approximately two $80$ MHz stop bands within an overall band of $1$ GHz ($W=2$ GHz).  
Fig.~\ref{fig:cognitive_radio_ex} shows two noncompressive estimates and compares them to a Welch estimate of the same resolution and calculated using Nyquist samples.
The left plot sets $q=25$ and $L=64$ and the right plot sets $q=50$ and $L=128$, meaning that the resolution of left hand estimate is half the resolution of the right hand estimate (half as fine) and that the sampling rate per channel on the left is twice that of the right (31.25 MHz vs 15.625 MHz)
In each case, however, the overall average system sampling rate is the same (781.25 MHz).
The noncompressive estimate on the left is formed with 4096 MC samples per channel, the one on the right with 2096 samples. 
Both plots suggest that spectral holes could be detected simply by thresholding the estimates and hence noncompressive estimators may provide a way to monitor large bandwidths at low sampling rates.

\section{Conclusion}\label{sect:concl}
In this paper, we derived and analyzed a consistent, MC based PSD estimator that produces both compressive and noncompressive estimates at sub-Nyquist sampling rates.
The estimator does not estimate the power spectrum on a discrete grid of frequencies but instead computes average power within a given set of subbands.
Compressive estimates leverage the sparsity of the spectrum and exhibit better tradeoffs among the estimator's resolution, complexity, and average sampling rate compared to noncompressive estimates.
Given suitable sampling patterns, both compressive and noncompressive estimates can be recovered using NNLS, thus avoiding the overhead of computing the estimates in different ways.
The estimator is consistent and has wide applicability; it is especially attractive in wideband applications where high sampling rates are costly or difficult to implement.

\section{Appendix}\label{appendix}
To show that $\widehat{\B{v}}_{NC}$ is asymptotically efficient, we show that it is a maximum likelihood estimator (MLE) because MLEs are known to be asymptotically efficient~\cite[p. 472]{castella_berger2002}.

Define $\B{z}(n)$ to be the $n$th snapshot of samples across all the channels of a MC sampler:
\begin{equation}
\B{z}(n)\triangleq [z_1(n), z_2(n), \dotsc, z_q(n)]^{T}.
\end{equation}
The entries of $\B{z}(n)$ are the nonuniform samples collected within one period ($L/W$ seconds) of the MC sampling sequence.   
The upper triangular portion of the sample correlation matrix $\B{S}$ are the elements of $\widehat{\B{u}}$:
\begin{align}
\B{S} &\triangleq \frac{1}{N} \sum_{n=0}^{N-1} \B{z}(n)\B{z}(n)^T\\
&=\begin{bmatrix}
\widehat{r}_{z_1z_1}^N(0) & \widehat{r}_{z_1z_2}^N(0) & \hdotsfor[2]{2} & \widehat{r}_{z_1z_q}^N(0) \\
\widehat{r}_{z_2z_1}^N(0) & \widehat{r}_{z_2z_2}^N(0) & \hdotsfor[2]{2}  & \vdots \\
\vdots & &  \ddots & & \\
\widehat{r}_{z_qz_1}^N(0) & \hdotsfor[2]{3} & \widehat{r}_{z_qz_q}^N(0)
\end{bmatrix}.
\end{align}

Suppose $x(t)$ is zero-mean WSS Gaussian random process.
Then each snapshot $\B{z}(n)$ is an i.i.d. realization of a jointly Gaussian random vector with correlation matrix $\B{R}$.
Taking $\B{R}$ to be the parameter of interest, the log-likelihood function given the data $\{\B{z}(n)\}_{n=0}^{N-1}$ is~\cite{mullis_scharf1990,scharf1991}
\begin{equation}
L\big(\B{R};\{\B{z}(n)\}\big)=\frac{-qN}{2} \ln{2\pi} - \frac{N}{2} \ln{\det{(\B{R})}} - \frac{N}{2} \tr{\B{R}^{-1}\B{S}}.
\end{equation}
The maximum likelihood estimate of $\B{R}$ can be found by taking the gradient of $L\big(\B{R};\{\B{z}(n)\}\big)$ with respect to $\B{R}$ and setting the result equal to zero.
\begin{equation}
\frac{\partial L\big(\B{R};\{\B{z}(n)\}\big)}{\partial \B{R}} = -\frac{N}{2}(\B{R}^{-1})^T + \frac{N}{2}(\B{R}^{-1}\B{S}\B{R}^{-1})^T=0
\end{equation}
Solving for $\B{R}$ reveals that the maximum likelihood estimate of $\B{R}$ equals the sample correlation $\B{S}$, or in other words, the entries in $\widehat{\B{u}}$ are maximum likelihood estimates of $r_{z_az_b}(0)$ for all combinations of pairs $(a,b)$. 

Now since $\widehat{\B{v}}_{NC} = \B{\Psi}^{\dagger}\widehat{\B{u}}$, we have by the invariance property of maximum likelihood estimators~\cite{scharf1991,castella_berger2002} that $\widehat{\B{v}}_{NC}$ is a maximum likelihood estimate of $\B{v}$.
We therefore conclude that $\widehat{\B{v}}_{NC}$ is asymptotically efficient when $x(t)$ is a zero-mean WSS Gaussian random process.

\bibliographystyle{IEEEbib}
\bibliography{lexa}

\end{document}